\documentclass[twocolumn]{aastex63}

\usepackage{soul, color, xcolor,times}
\usepackage{amsmath}
\usepackage{natbib}

\definecolor{emerald}{rgb}{0.1,0.5,0.3}

\newcommand{\sub}{\textsubscript}
\newcommand{\ts}{\textsuperscript}
\usepackage{enumitem}
\graphicspath{{./}{figures/}}

\shorttitle{Disturbances in post-starburst galaxies seen with \textit{HST}.}
\shortauthors{Sazonova et al.}

\begin{document}


\title{Are all post-starbursts mergers? \textit{HST} reveals hidden disturbances in the majority of PSBs.}


\author[0000-0001-6245-5121]{Elizaveta Sazonova}
\affiliation{Johns Hopkins University, Department of Physics and Astronomy, Baltimore, MD 21218, USA}

\author[0000-0002-4261-2326]{Katherine Alatalo}
\affiliation{Space Telescope Science Institute, 3700 San Martin Dr, Baltimore, MD 21218, USA}
\affiliation{Johns Hopkins University, Department of Physics and Astronomy, Baltimore, MD 21218, USA}

\author[0000-0001-7883-8434]{Kate Rowlands}
\affiliation{AURA for ESA, Space Telescope Science Institute, 3700 San Martin Dr, Baltimore, MD 21218, USA}
\affiliation{Johns Hopkins University, Department of Physics and Astronomy, Baltimore, MD 21218, USA}

\author[0000-0003-2823-360X]{Susana E. Deustua}
\affiliation{Space Telescope Science Institute, 3700 San Martin Dr, Baltimore, MD 21218, USA}

\author[0000-0002-4235-7337]{K. Decker French}
\affiliation{Department of Astronomy, University of Illinois Urbana-Champaign, Urbana, IL 61801, USA}

\author[0000-0001-6670-6370]{Timothy Heckman}
\affiliation{Johns Hopkins University, Department of Physics and Astronomy, Baltimore, MD 21218, USA}

\author[0000-0002-3249-8224]{Lauranne Lanz}
\affiliation{The College of New Jersey, Ewing, NJ 08618, USA}

\author[0000-0002-9471-5423]{Ute Lisenfeld}
\affiliation{Dpto. de Física Teórica y del Cosmos, Campus de Fuentenueva, Edificio Mecenas, Universidad de Granada, E-18071 Granada, Spain}
\affiliation{Instituto Carlos I de Física Téorica y Computacional, Facultad de Ciencias, 18071 Granada, Spain}

\author[0000-0002-0696-6952]{Yuanze Luo}
\affiliation{Johns Hopkins University, Department of Physics and Astronomy, Baltimore, MD 21218, USA}

\author[0000-0001-7421-2944]{Anne Medling}
\affiliation{Ritter Astrophysical Research Center, University of Toledo, Toledo, OH 43606, USA}
\affiliation{ARC Centre of Excellence for All Sky Astrophysics in 3 Dimensions (ASTRO 3D)}

\author[0000-0003-1991-370X]{Kristina Nyland}
\affiliation{National Research Council, resident at the U.S. Naval Research Laboratory, 4555 Overlook Ave SW, Washington, DC 20375, USA}

\author[0000-0003-3191-9039]{Justin A. Otter}
\affiliation{Johns Hopkins University, Department of Physics and Astronomy, Baltimore, MD 21218, USA}

\author[0000-0003-4030-3455]{Andreea O. Petric}
\affiliation{Space Telescope Science Institute, 3700 San Martin Dr, Baltimore, MD 21218, USA}

\author[0000-0002-4226-304X]{Gregory F. Snyder}
\affiliation{Space Telescope Science Institute, 3700 San Martin Dr, Baltimore, MD 21218, USA}

\author[0000-0002-0745-9792]{Claudia Megan Urry}
\affiliation{11 Yale Center for Astronomy \& Astrophysics, 46 Hillhouse Avenue, New Haven, CT 06511, USA}
\affiliation{12 Department of Physics, Yale University, P.O. Box 208120, New Haven, CT 06520-8120, USA}

\begin{abstract}



How do galaxies transform from blue, star-forming spirals to red, quiescent early-type galaxies? To answer this question, we analyzed a set of 26 gas-rich, shocked post-starburst galaxies with \textit{Hubble Space Telescope} (\textit{HST}) imaging in \textit{B}, \textit{I}, and \textit{H} bands, and Sloan Digital Sky Survey (SDSS) \textit{i}-band imaging of similar depth but lower resolution. We found that post-starbursts in our sample have intermediate morphologies between disk- and bulge-dominated (Sérsic $n=1.7^{+0.3}_{-0.0}$) and have red bulges, likely due to dust obscuration in the cores.

Majority of galaxies in our sample are more morphologically disturbed than regular galaxies (88\%, corresponding to $>$3$\sigma$ significance) when observed with \textit{HST}, with asymmetry and Sérsic residual flux fraction being the most successful measures of disturbance. Most disturbances are undetected at the lower resolution of SDSS imaging. Although $\sim$27\% galaxies are clear merger remnants, we found that disturbances in another $\sim$30\% of the sample are internal, caused by small-scale perturbations or dust substructures rather than tidal features, and require high-resolution imaging to detect. We found a 2.8$\sigma$ evidence that asymmetry features fade on timescales $\sim 200$ Myr, and may vanish entirely after $\sim$750 Myr, so we do not rule out a possible merger origin of all post-starbursts given that asymmetric features may have already faded. This work highlights the importance of small-scale disturbances, detected only in high-resolution imaging, in understanding structural evolution of transitioning galaxies.

\vspace{3em}
\end{abstract}

\section{Introduction} \label{sec:intro}

There is a strong bimodality amongst galaxies in the local Universe: the majority of galaxies belong either to the blue star-forming cloud or the red quiescent sequence \citep{Baldry2004}. Over cosmic time, the number density of galaxies increased on the red sequence and decreased in the blue cloud, showing that star-forming galaxies quench their star formation and transition to the red sequence \citep{Bell2004, Arnouts2007, Faber2007}. Transitioning galaxies, found in the ``green valley", are relatively rare \citep{Bell2003}, and the majority of ``green valley" galaxies are quenching slowly by building up the central bulge and gradually exhausting their star forming fuel \citep[e.g.,][]{Noeske2007, Schawinski2014}. However, some galaxies quench much more rapidly, on timescales of $\leq$1 Gyr, and therefore are much rarer \citep[e.g.,][]{Goto2005}. It is difficult to find a large enough population of rapidly transitioning galaxies to study, and the mechanism behind this rapid quenching is still one of the biggest unresolved questions in galaxy evolution.


Another important bimodality in the local galaxy population is that of the galaxy structure. Blue galaxies are predominantly disks with spiral arms, while red galaxies are dominated either partially or entirely by a central bulge, which is a spheroidal component \citep{Strateva2001, Schawinski2014}. Blue spheroids and red disks are extremely uncommon in the local Universe \citep{Masters2010, Barro2013}. This bimodality provides insight into the galaxies' evolutionary pathway: during the short transition period, the galaxies must both stop forming stars and change their morphology. However, the relative timescales are still uncertain: does morphology of these rapidly transitioning galaxies change prior to, concurrently with, or after the quenching of star formation?

\subsection{Post-starburst galaxies}
To answer these questions, it is essential to study galaxies that are currently rapidly transitioning. In recent years, large surveys such as the Sloan Digital Sky Survey \citep[SDSS; e.g.,][]{dr2, York2000, Strauss2002, sdss4} have provided imaging and spectral data for millions of galaxies. Among these, a rare class of galaxies called ``post-starbursts" (PSBs) have been identified, whose star formation quenched in the past $\sim$Gyr \citep[e.g.,][]{Dressler1983, Zabludoff1996, Goto2005}. Although some PSBs rejuvenate and continue to form stars, most do not; therefore they are the ideal sample to study rapid galaxy quenching \citep{Young2014}.

\begin{figure*}
    \centering
    \includegraphics[width=\linewidth]{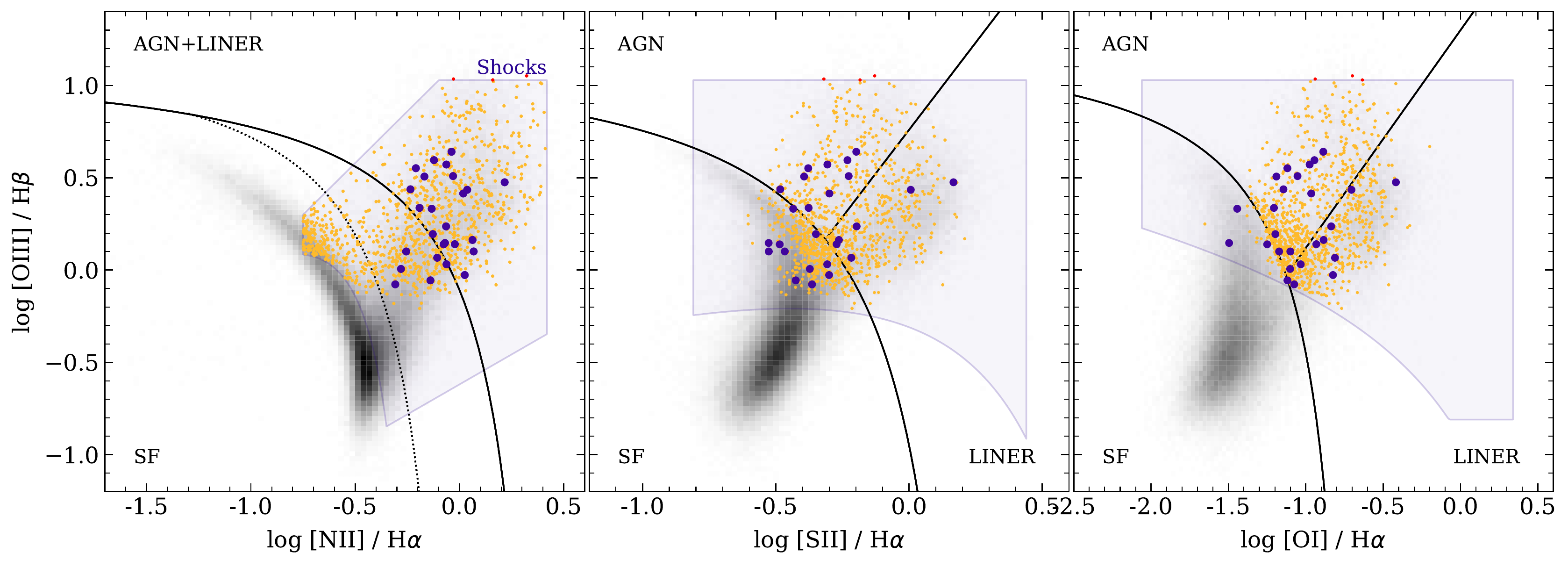}
    \caption{ The line diagnostic diagrams \citep{Baldwin1981, Veilleux1987} used to determine the ionization mechanism with spectroscopic data for [\ion{O}{3}]/H$\beta$, [\ion{N}{2}]/H$\alpha$, [\ion{S}{2}]/H$\alpha$ (using the [\ion{S}{2}] doublet) and [\ion{O}{1}]/H$\alpha$ emission lines. Three data sets are plotted: ELG sample (\textbf{grey}), SPOGs (\textbf{orange)}, and HST-SPOGs (\textbf{purple}). Regions of star formation (SF), AGN, and LINER emission are separated by black solid lines \citep{Kauffmann2003, Kewley2006}. The purple outline shows the shock ionization boundaries from \cite{spogs}. 3 galaxies shown in red are SPOGs that have ionization above the shock boundary and are therefore excluded from our sample.} 
    \label{fig:bpt}
\end{figure*}

Historically, PSBs have been identified by a lack of ionized emission lines caused by short-lived massive stars and strong H$\delta$ absorption from intermediately-age A stars \citep[``K+A" or ``E+A" galaxies;][]{Quintero2004, Goto2005, Goto2007}. The morphology of E+As varies from intermediate to early-type, indicating they are likely transitioning galaxies \citep{Quintero2004, Tran2004, Blake2004, Yang2008, Pawlik2018}. However, this selection is biased against other energetic processes associated with quenching that could result in strong emission lines, such as shocks induced by stellar winds, Active Galactic Nuclei (AGN), or low-ionization nuclear emission line regions (LINER). To account for this, several new methods have been proposed that do not exclude ionized emission \citep{Wild2007, Yesuf2014}. One such method selects a set of Shocked POst-starburst Galaxies \citep[SPOGs;][]{spogs}, by allowing ionized emission consistent with shocks, AGN, and LINERs rather than purely star formation \citep{Kauffmann2003, Kewley2006}, as shown in Fig. \ref{fig:bpt}. SPOGs trace, on average, a younger population of PSBs than E+As, so they are a powerful set of galaxies to study the early phase of the transition \citep{spogs}.


\subsection{Quenching Mechanisms}
An array of quenching mechanisms capable of quickly shutting down star formation have been proposed, but it is unclear which are dominant for which galaxies, and it is likely they act in tandem rather than on their own.  Quenching models act by either removing or heating the star-forming molecular gas to prevent it from collapsing into stars. PSBs still host large reservoirs of molecular gas \citep{French2015, Rowlands2015} but do not form stars, while red sequence galaxies are devoid of gas \citep{Young2011, Crocker2011}. Therefore, gas suppression is important in triggering the PSB phase, while gas removal is necessary to complete the transition to a quiescent galaxy.

Among others, supernovae winds \citep[e.g.,][]{Kaviraj2007}, AGN feedback \citep[e.g.,] []{Cicone2014, Baron2019}, shocks \citep[e.g.,][]{Alatalo2015}, turbulence \citep[e.g.,][]{Lanz2016,Smercina2018}, and ram pressure stripping \citep{Gunn1972} in galaxy clusters have all been invoked to either remove gas or prevent it from forming stars. However, these processes simply affect the gas supply and do not transform the galaxy morphology on their own. Since star-forming and quiescent galaxies have starkly different structures, one or more other mechanisms must act alongside gas suppression to change the structure of a transitioning galaxy. Therefore, to fully understand the quenching pathway(s) of post-starbursts, it is essential to study their morphology.


Galaxy mergers are an effective way to disturb the galactic disk and build up a central bulge, changing the galaxy morphology \citep[e.g.,][]{Toomre1972, Hopkins2006, Snyder2011}, and have been shown an important formation mechanism for simulated post-starbursts \citep{Zheng2020,Lotz2020}. They can drive shocks across the interstellar medium (ISM) and trigger a central starburst or an AGN, leading to strong supernovae- and/or AGN-driven winds \citep{Bekki2005}. Therefore, mergers are an attractive candidate for the primary quenching trigger. Recent studies find that about 50\% of PSBs show merger signatures, although the merger fraction varies from 15\% to 70\% \citep{Zabludoff1996, Goto2005, Yang2008, Pracy2009, Pawlik2016}. Although the merger fraction of PSBs is larger relative to the average fraction in the field \citep[1.5\% - 5\%, ][]{Darg2010}, roughly half of PSBs do not appear to be merger remnants. We then face a question: what triggered a starburst in the remaining, non-merger PSBs?

The remaining galaxies may have had their morphology transformed via less violent events: minor mergers \citep[e.g.,][]{Bournaud2007,Jackson2019}, tidal torques \citep[e.g.,][]{Moore1996}, disk instabilities \citep[e.g.,][]{Dekel2013}, or a gradual build-up of a central bulge leading to morphological quenching \citep[e.g.,][]{Martig2013,Gensior2020}.

On the other hand, it is possible that the majority of PSBs experienced a merger event that remains undetected. Simulations show that merger signatures fade on the timescale of 100-500 Myr \citep{Lotz2008, Snyder2015b, Pawlik2018}, depending on the parameters of the merger. Traditionally, mergers are identified in observational data via asymmetry in the galaxy light profile \citep[e.g.,][]{Conselice2003, Pawlik2016}. Longer-lasting merger signatures such as shells are symmetric and therefore would be undetected. The seeing limit of ground-based surveys also limits our ability to detect small-scale disturbances below the resolution limit. 


To determine whether galaxy mergers are common in PSBs, we obtained high-resolution optical imaging of a set of PSBs with the \textit{Hubble Space Telescope} (HST), with comparable depth to existing lower-resolution SDSS imaging. We observed a sample of shocked PSBs that contain molecular gas, tracing a relatively young PSB phase. With careful modelling of galaxy light using \textsc{GALFIT} \citep{galfit} as well as non-parametric analysis using \textsc{statmorph} \citep{statmorph}, we determined whether these quckly transitioning galaxies are consistent with spheroids or disks, and whether they show more disturbance than regular star-forming and quiescent galaxies.

The paper is arranged as follows. In Section 2, we present the data sample used in this study, including the sub-selection of our subsample of SPOGs and the comparison sample. In Section 3, we describe the imaging data reduction and preparation. In Section 4, we explain the image processing tools used to compute morphology of our sample. In Section 5, we investigate the physical properties of our sample and their effects on morphology. In Section 6, we compare the morphology of our sample of PSBs to the comparison sample of star-forming and quiescent galaxies. Finally in Section 7, we discuss the implications of our results on the likelihood of the merger origin of PSBs. 

Throughout this work, we used \textit{Wilkinson Microwave Anisotropy Probe 9} (WMAP 9) cosmology with $(\Omega_\Lambda, \Omega_M, h) = (0.713, 0.287, 0.693)$ \citep{wmap9}.

\vspace{-2mm}
\section{Data Selection}\label{sec:data}

\begin{figure*}
    \centering
    \includegraphics[width=\linewidth]{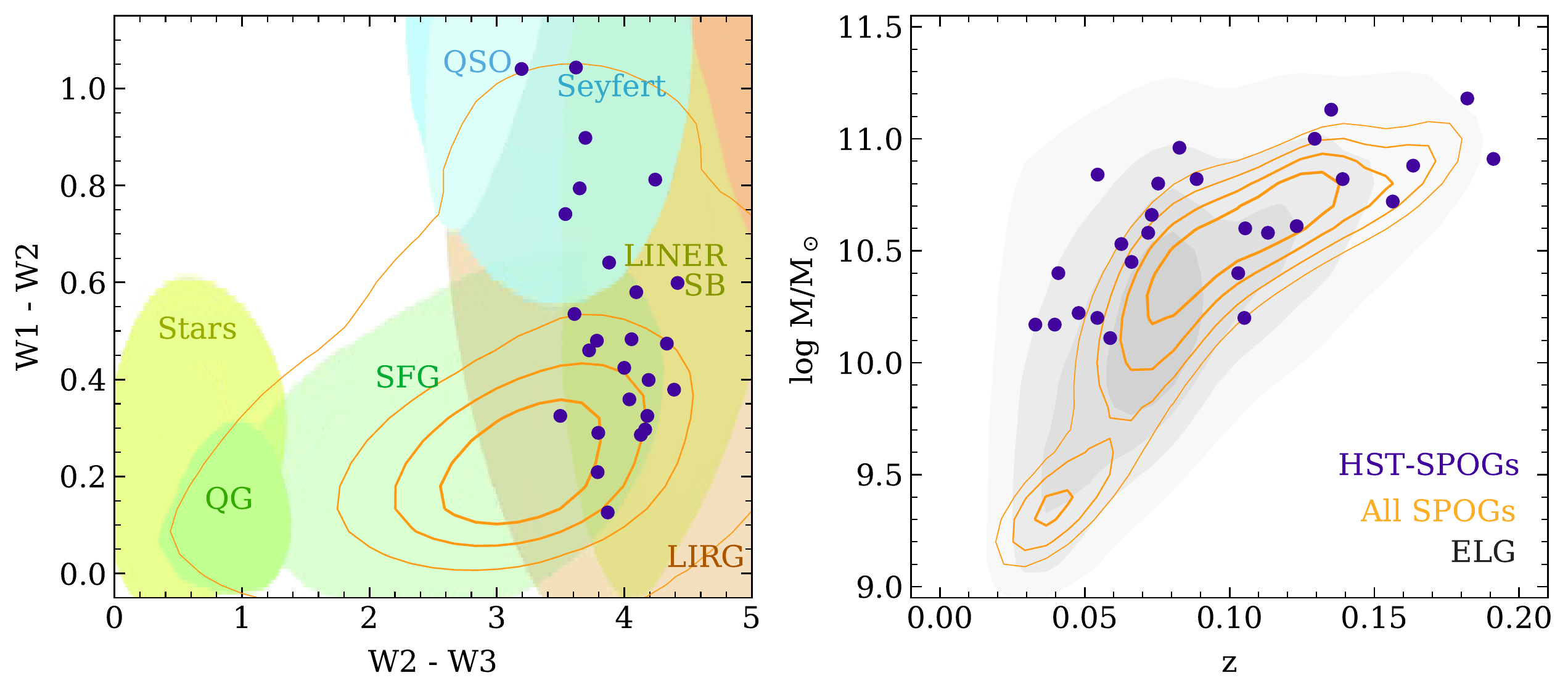}
    \caption{\textbf{Left:} redshift and stellar mass distributions of the ELG (\textbf{grey}), SPOG (\textbf{orange}), and HST-SPOG (\textbf{purple}) samples. HST-SPOGs are, on average, slightly more massive than the parent SPOG sample. \textbf{Right:} WISE color distributions of the SPOG (\textbf{orange}) and HST-SPOG (purple) samples. Typical regions for stars, quiescent galaxies (QGs), star-forming galaxies (SFGs), luminous IR galaxies (LIRGs), LINERs, starbursts (SBs), Seyfert galaxies, and quasars (QSOs) from \cite{Wright2010} are overplotted. HST-SPOGs generally lie in the LINER/starburst regions.} 
    \label{fig:mass-z}
\end{figure*}

\subsection{Shocked Post-Starburst Galaxies}

\begin{deluxetable*}{ccccccccccccc}
\label{tab:data_summary}
\tablecaption{Summary of HST-SPOG sample properties and imaging data used in this work.}
\tablehead{IAU Name & ID & R.A. & Dec. & Redshift & log M$_\star/$M$_\odot$ & \textit{NUV}-\textit{r} & $\tau_{\textrm{SB}}$ & $f_{\textrm{gas}}$ & \multicolumn{4}{c}{1$\sigma$ image depth [mag/arcsec\ts{2}]} \\ 
& & [hms] & [dms] & & & [mag] & [Myr] & & F438W & F814W & F160W & \textit{i}
}
\colnumbers
\startdata 
\object{J000318+004844} & 1 & 00:03:18.21 & +00:48:44 & 0.139 & 10.8$^{+0.1}_{-0.1}$ & 3.46 &153$^{+421}_{-48}$ & 0.24 & 23.4 & 25.1 & 23.4 & 23.7 \\
\object{J001145-005431} & 4 & 00:11:45.21 & -00:54:30 & 0.048 & 10.2$^{+0.1}_{-0.1}$ & 5.07 &692$^{+268}_{-106}$ & 0.06 & 22.4 & 24.6 & 23.4 & 23.6 \\
\object{J011957+133431} & 24 & 01:19:56.76 & +13:34:31 & 0.191 & 10.9$^{+0.1}_{-0.1}$ & 3.04 &75$^{+37}_{-23}$ & 0.25 & 23.6 & 24.4 & 23.3 & 23.3 \\
\object{J080400+253051} & 77 & 08:03:59.62 & +25:30:51 & 0.135 & 11.1$^{+0.1}_{-0.2}$ & \nodata & 109$^{+104}_{-119}$ & 0.11 & 23.4 & 24.9 & 23.4 & 23.4 \\
\object{J080724+200608} & 81 & 08:07:24.46 & +20:06:08 & 0.066 & 10.4$^{+0.1}_{-0.1}$ & 4.60 &340$^{+53}_{-47}$ & 0.12 & 24.0 & 23.4 & 23.1 & 22.4 \\
\object{J081603+193643} & 98 & 08:16:03.15 & +19:36:43 & 0.113 & 10.6$^{+0.1}_{-0.1}$ & 2.79 &61$^{+32}_{-16}$ & 0.17 & 23.5 & 23.5 & 23.4 & 24.8 \\
\object{J084545+200610} & 142 & 08:45:45.38 & +20:06:10 & 0.123 & 10.6$^{+0.1}_{-0.1}$ & \nodata &78$^{+83}_{-51}$ & 0.27 & 23.1 & 23.4 & 23.4 & 22.2 \\
\object{J085357+031034} & 157 & 08:53:56.81 & +03:10:33 & 0.129 & 11.0$^{+0.1}_{-0.1}$ & 4.14 &52$^{+198}_{-238}$ & 0.20 & 23.4 & 23.4 & 23.8 & 22.4 \\
\object{J085943+100644} & 169 & 08:59:42.62 & +10:06:43 & 0.054 & 10.8$^{+0.1}_{-0.1}$ & 4.40 &306$^{+67}_{-43}$ & 0.07 & 23.6 & 23.6 & 23.4 & 24.9 \\
\object{J091407+375310} & 186 & 09:14:07.22 & +37:53:09 & 0.072 & 10.6$^{+0.1}_{-0.1}$ & 3.58 &127$^{+87}_{-82}$ & 0.28 & 23.5 & 23.4 & 23.4 & 22.5 \\
\object{J091850+420044} & 191 & 09:18:49.99 & +42:00:43 & 0.041 & 10.4$^{+0.0}_{-0.1}$ & 4.99 &490$^{+99}_{-74}$ & 0.03 & 23.4 & 23.4 & 23.4 & 22.4 \\
\object{J092518+062334} & 200 & 09:25:18.31 & +06:23:34 & 0.075 & 10.8$^{+0.1}_{-0.2}$ & 3.75 &171$^{+146}_{-76}$ & 0.10 & 23.4 & 23.4 & 23.4 & 22.5 \\
\object{J092820+074159} & 209 & 09:28:19.53 & +07:41:58 & 0.105 & 10.2$^{+0.2}_{-0.1}$ & 2.25 &-4$^{+4}_{-5}$ & 0.30 & 23.4 & 23.5 & 23.4 & 23.4 \\
\object{J093820+181953} & 224 & 09:38:19.87 & +18:19:52 & 0.089 & 10.8$^{+0.0}_{-0.2}$ & 4.61 &197$^{+55}_{-109}$ & 0.27 & 23.4 & 23.9 & 23.2 & 22.2 \\
\object{J095750-001239} & 253 & 09:57:49.54 & -00:12:39 & 0.033 & 10.2$^{+0.1}_{-0.2}$ & 4.48 &381$^{+59}_{-59}$ & 0.07 & 23.3 & 23.3 & 23.4 & 24.8 \\
\object{J100829+191620} & 268 & 10:08:28.73 & +19:16:19 & 0.182 & 11.2$^{+0.1}_{-0.1}$ & 2.87 &166$^{+339}_{-108}$ & 0.20 & 23.2 & 23.3 & 23.3 & 22.3 \\
\object{J100848+512353} & 270 & 10:08:47.69 & +51:23:52 & 0.156 & 10.7$^{+0.1}_{-0.1}$ & 2.34 &-94$^{+147}_{-162}$ & 0.33 & 23.4 & 23.4 & 23.3 & 22.5 \\
\object{J102653+434008} & 305 & 10:26:53.35 & +43:40:08 & 0.105 & 10.6$^{+0.0}_{-0.0}$ & \nodata & 60$^{+18}_{-25}$ & 0.24 & 23.6 & 23.4 & 23.4 & 22.4 \\
\object{J102826+573609} & 308 & 10:28:25.80 & +57:36:09 & 0.073 & 10.7$^{+0.1}_{-0.1}$ & 3.23 &-13$^{+26}_{-148}$ & 0.34 & 23.4 & 23.4 & 23.4 & 22.5 \\
\object{J103135+054057} & 322 & 10:31:34.85 & +05:40:57 & 0.163 & 10.9$^{+0.1}_{-0.1}$ & 4.02 &142$^{+42}_{-38}$ & 0.33 & 23.4 & 24.7 & 23.4 & 23.3 \\
\object{J105751+055447} & 365 & 10:57:51.07 & +05:54:46 & 0.054 & 10.2$^{+0.1}_{-0.1}$ & 3.24 &23$^{+56}_{-21}$ & 0.17 & 23.4 & 23.4 & 23.4 & 24.7 \\
\object{J112619+191329} & 437 & 11:26:19.44 & +19:13:29 & 0.103 & 10.4$^{+0.0}_{-0.1}$ & 2.70 &-84$^{+25}_{-125}$ & 0.45 & 23.4 & 23.4 & 23.5 & 22.2 \\
\object{J124822+551452} & 619 & 12:48:22.17 & +55:14:52 & 0.083 & 11.0$^{+0.1}_{-0.1}$ & \nodata & 433$^{+531}_{-270}$ & 0.13 & 23.5 & 23.5 & 23.5 & 22.4 \\
\object{J133953+442237} & 711 & 13:39:53.19 & +44:22:36 & 0.063 & 10.5$^{+0.2}_{-0.1}$ & 5.10 &211$^{+418}_{-166}$ & 0.24 & 23.4 & 23.4 & 23.4 & 22.3 \\
\object{J150619+080642} & 862 & 15:06:19.17 & +08:06:42 & 0.040 & 10.2$^{+0.1}_{-0.1}$ & \nodata &4$^{+8}_{-10}$ & 0.07 & 23.5 & 24.9 & 23.8 & 23.5 \\
\object{J164504+304802} & 1014 & 16:45:03.79 & +30:48:02 & 0.059 & 10.1$^{+0.0}_{-0.0}$ & 4.74 &258$^{+498}_{-203}$ & 0.19 & 23.4 & 22.3 & 23.4 & 23.4 \\
\enddata 
\tablecomments{(1): IAU name. (2): ID in the SPOG sample catalog from \citet[Tab. 1]{spogs}. (3) and (4): J2000 coordinates. (5): SDSS spectroscopic redshift. (6): stellar mass from  \cite{Chang2015} catalog. (7): \textit{NUV}-\textit{r} color, where the galaxy is detected with \textit{GALEX}. (8): post-starburst age from \citet[Tab. 2]{French2018}. (9): molecular gas fraction from \citet[Tab. 3]{spogs2}. (10), (11), (12), (13): 1$\sigma$ sky background flux of \textit{HST} F438W, F814W, F160W, and SDSS \textit{i}-band imaging respectively in AB mag/arcsec\ts{2}.}
\end{deluxetable*}

We analyzed a subset of SPOGs \citep{spogs} in this study. SPOGs were identified using the Oh-Sarzi-Schawinski-Yi catalog  \citep[OSSY;][]{Oh2011} of emission and absorption features in galaxies from the  SDSS Data Release 7 \citep[DR7;][]{Abazajian2009} spectroscopy. 

First, a parent emission line galaxy (ELG) sub-sample was selected from the OSSY catalog to ensure high signal-to-noise (S/N) data in the spectral continuum and for the H$\beta$, [\ion{O}{3}], H$\alpha$, [\ion{N}{2}], [\ion{S}{2}] and [\ion{O}{1}] narrow lines. ELG galaxies must have stellar continuum $S/N > 10$ in the 4500\AA\ -- 7000\AA\, wavelength range and line amplitude-to-noise ratios $A/N > 1$ for each line. Additionally, the ELG sample required that the ratio of the fit residuals to statistical noise ($N_\sigma$) is less than 3$\sigma$ and 5$\sigma$ away from the OSSY sample average for continuum and line fits respectively. The ELG sample consists of 159,387 galaxies (24$\pm$0.05\% of the OSSY catalog). SPOGs were selected from the ELG sample using the following criteria:

\begin{enumerate}[itemsep=0mm]
\item Strong Balmer absorption characteristic of post-starbursts \citep[Lick H$\delta_A$ index $>$ 5\AA,][]{Goto2007};
    
    \item Emission line ratios inside the shocked region on each diagnostic diagram in Fig. \ref{fig:bpt} \citep{Baldwin1981,Veilleux1987,Kauffmann2003, Kewley2006, spogs};
    
    \item Emission line ratios outside of the star formation or composite region on at least one diagram in Fig. \ref{fig:bpt}.
    
\end{enumerate}

This method selects galaxies that host a population of intermediately-aged A-stars, which give strong Balmer absorption, but not short-lived massive stars, which would ionize the gas. The requirement to have strong emission lines, unassociated with star formation, selects galaxies with ongoing energy injection from shocks, AGN winds, or LINER emission. The SPOGs sample contains 1067 galaxies. Three galaxies from the parent SPOG sample (shown in red in Fig. \ref{fig:bpt}) have $\log$([\ion{O}{3}]/H$\beta$) slightly above the shock region cut-off, so we excluded them from our sample.

\cite{spogs2} described the CO(1-0) emission in 53 SPOGs using Institut de Radioastronomie Millimétrique (IRAM) 30 m single dish and the Combined Array for Research for Millimeter Astronomy (\textit{CARMA}) interferometer. This set of galaxies was selected from a subsample of SPOGs that had a detectable 22$\mu$m emission in the \textit{Wide-Field Infrared Survey} (\textit{WISE}) all-sky survey data. Of those 53, 47 galaxies were detected to have CO(1-0) emission and hence molecular gas; approximately 40\% of the observed galaxies were visually disturbed in SDSS imaging.


To further study the morphology and dust content of the 47 CO-detected SPOGs, we have obtained \textit{HST} imaging for 26 galaxies as part of the snapshot program (Proposal 14649, PI: Alatalo).  These galaxies (hereafter HST-SPOGs) were observed using \textit{HST} Wide-Field Camera 3 (WFC3) with F438W (\textit{B}), F814W (\textit{I}) and F160W (\textit{H}) filters in order to capture the dust extinction and study morphology at high resolution.

This work focuses on a morphological analysis of the 26 HST-SPOGs and a comparison to the morphological parameters obtained from existing SDSS imaging of the same galaxies with comparable depth. We use SDSS \textit{g} and \textit{i} filters as the closest matched filters to \textit{B} and \textit{I} \textit{HST} observations, respectively. 

We obtained spectroscopic redshifts for all HST-SPOGs from the NYU Value-Added Catalog \citep[NYU-VAGC;][]{Blanton2005}. We used a stellar mass catalog derived from the spectral energy distribution (SED) fitting of NYU-VAGC galaxies using SDSS and \textit{WISE} photometric data from \cite{Chang2015}. This catalog contained only 148,921 out of 155,364 ELG galaxies, including 975 out of 1067 SPOGs and 25 out of 26 HST-SPOGs. For one HST-SPOG that was missing a stellar mass measurement from \cite{Chang2015}, we used the stellar mass from MPA-JHU catalog \citep{mpa}. Although MPA-JHU stellar mass estimates exist for all SPOGs, we opted for the \cite{Chang2015} catalog because they include IR \textit{WISE} data, making the mass estimates more robust. There is no systematic offset between the two catalogs, and the scatter is small. We used post-starburst age estimates that were measured in \cite{French2018} by fitting galaxy SEDs with stellar population synthesis models to obtain star formation histories. 

The diagnostic emission line ratios for the HST-SPOGs are overplotted in purple in Fig. \ref{fig:bpt}, showing that the ionization mechanisms in HST-SPOGs are consistent with the parent SPOGs sample. Fig. \ref{fig:mass-z} (left) shows the distribution of \textit{WISE} colors for SPOGs (orange) and HST-SPOGs (purple). Overplotted is the adaptation of different color-color regions corresponding to different types of galaxies from \cite{Wright2010}. HST-SPOGs occupy primarily the starburst/LINER region, which is consistent with the fact that they all lie in the AGN/LINER or composite regions on the BPT diagram. Finally, Fig. \ref{fig:mass-z} (right) shows the distributions of stellar masses and redshifts for the ELG (grey), SPOG (orange) and HST-SPOG (purple) samples. SPOGs are found at slightly higher redshifts than the ELG sample, and HST-SPOGs match the SPOG redshift distribution well. However, HST-SPOGs appear more massive on average than the SPOG sample.

A summary of our sample, including galaxies' redshift, stellar mass, \textit{NUV}-\textit{r} color, post-starburst age, gas fraction $f_{\textrm{gas}}$, and $1\sigma$ image depth in each of the filters used, is given in Tab.  \ref{tab:data_summary}.

\vspace{-0.5em}
\subsection{Comparison Sample}\label{sec:control}

Measurements of galaxy morphology depend strongly on imaging properties \citep[such as image wavelength, depth, and spatial resolution, e.g.,][]{Lotz2004}. Before performing any analysis on a specific population of galaxies, it is essential to calibrate the morphology measurements for the given type of imaging against normal, secularly evolving galaxies, so that we are able to detect irregular morphology. Therefore, we constructed a \textit{comparison} sample and used it to define a set of \textit{normal} morphology measurements for star-forming and quiescent galaxies. We then compared the morphology of HST-SPOGs against that of the sample of regular galaxies to detect the irregularities in their morphology. 

We calibrated the morphology of HST-SPOGs against two samples: star-forming and quiescent galaxies. Stellar mass strongly affects the evolution of the galaxy \citep[e.g.,][]{Kauffmann2003, Baldry2004}, so it is crucial to use a comparison sample of a matching stellar mass. Moreover, galaxies with higher redshift or smaller mass are fainter and smaller. Since morphological measurements depend on image resolution and depth \citep[e.g.,][]{Lotz2004},  measurements of smaller and fainter galaxies will systematically differ to the measurements of better-resolved ones. To account for these biases, we found one quiescent and one star-forming galaxy for each HST-SPOG that most optimally matches that HST-SPOG's mass and redshift.

The comparison sample selection faced a number of constraints: we required the comparison galaxies to have 1) \textit{HST} imaging to at least the depth of the snapshot program, 2) SDSS coverage, 3) spectroscopic redshift measurements, 4) stellar mass estimates, and 5) the observations (described below) necessary to select star-forming or quiescent galaxies.


To determine whether a galaxy is star-forming or quiescent, we chose to use the catalog of  \textit{Galaxy Evolution Explorer} (\textit{GALEX}) Near-Ultraviolet (\textit{NUV}) Data Release 7 (DR7) observations. The \textit{NUV}-\textit{r}  color is a powerful probe into the star formation history and can distinguish actively star-forming galaxies from quiescent ones more robustly than optical color diagnostics \citep{Kaviraj2007_nuv, Kaviraj2009}. \textit{GALEX} has a full-sky catalog, therefore the requirement of \textit{GALEX} data did not constrain our sample more than the requirement to have SDSS and \textit{HST} observations.

We used the same \textit{WISE}-SDSS photometric mass catalog \citep{Chang2015} and spectroscopic redshifts from NYU-VAGC \citep{Blanton2005} as for HST-SPOGs.

Finally, to find F814W \textit{HST} imaging of the required depth, we used the Hubble Source Catalog \citep[HSC, 3rd version;][]{hsc}. This is a catalog obtained from performing object detection and photometry on all publicly available \textit{HST} data in the Hubble Legacy Archive.

The procedure to obtain the comparison sample candidates was as follows:

\begin{enumerate}
    \item Select galaxies with  $9 < \log M_\star/M_\odot < 12$ and $z < 0.2$ from the \cite{Chang2015} catalog; resulting in 666,212 galaxies.
    \item Cross-match this selection with HSC to find \textit{HST} observations, using a coordinate search with a 1.5\arcsec  search radius; resulting in 7046 galaxies.
    \item Cross-match with SDSS Data Release 12 \citep[DR12;][]{Alam2015} using a 1.5\arcsec coordinate search; still leaving 7046 galaxies.
    \item SDSS DR12 photometric catalog has already been coordinate-matched to the \textit{GALEX} DR7 catalog with a 5\arcsec search radius with multiple matches for each object \citep{Budavari2009}. Select the closest \textit{GALEX} object to each SDSS object.
    \item Select objects with \textit{NUV} S/N $>$ 3, leaving 1669 galaxies.
\end{enumerate}
\begin{figure}
    \centering
    \includegraphics[width=\linewidth]{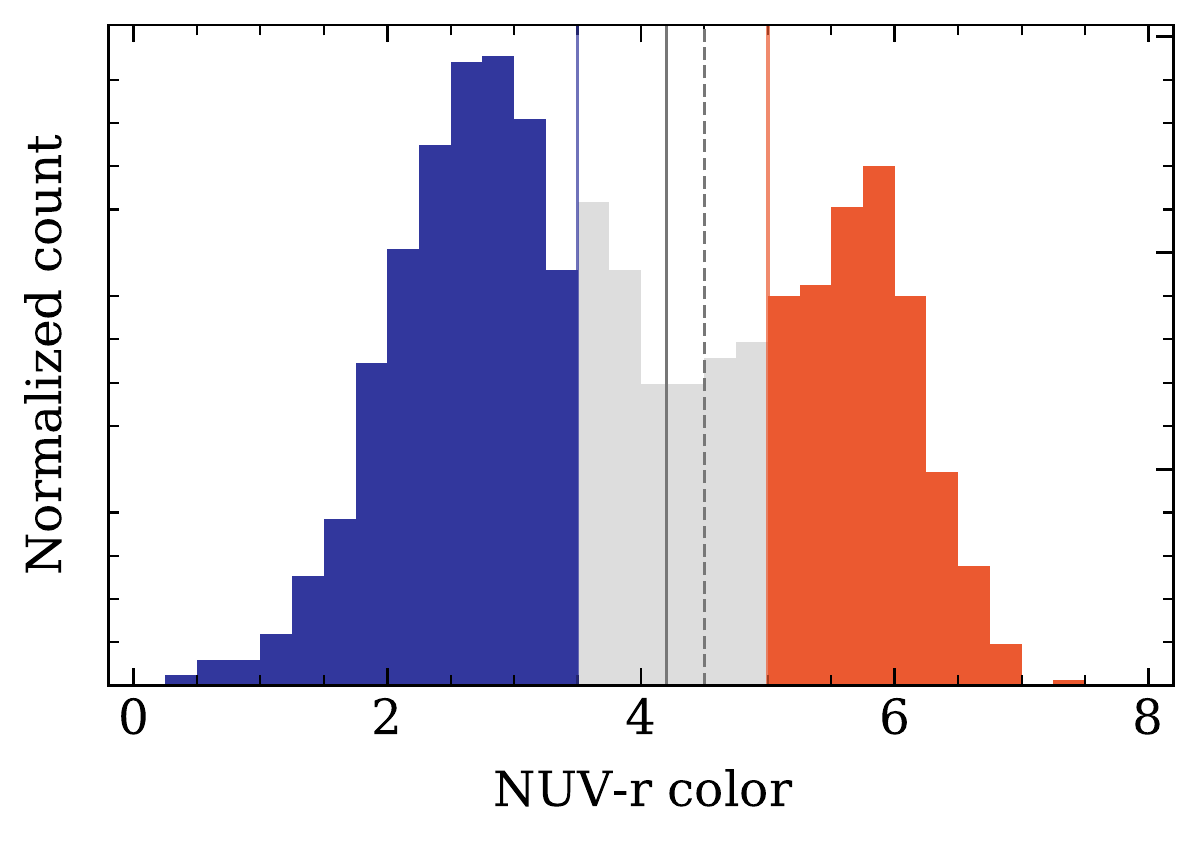}
    \caption{The distribution of \emph{GALEX} \textit{NUV} - SDSS \textit{r} colors of the 1,669 comparison sample candidates shows a clear bimodality. The approximate minimum of the distribution is at \textit{NUV}-\textit{r}  = 4.2 (solid grey line). Dashed grey line shows \textit{NUV}-\textit{r} $=$ 4.5, used in \cite{Kaviraj2007_nuv, Kaviraj2009}. The blue and the red regions indicate the cuts we used to select star-forming and quiescent galaxies, respectively.}
    \label{fig:nuv_cut}
\end{figure}

We then looked at the \textit{NUV}-\textit{r}  color distribution of the remaining 1669 comparison sample candidates, shown in Fig. \ref{fig:nuv_cut}. There is a clear bimodality around \textit{NUV}-\textit{r}  $\approx$ 4.3, consistent with the \textit{NUV}-\textit{r}  $\approx$ 4.5 cut used in \cite{Kaviraj2007_nuv} and \cite{Kaviraj2009} to distinguish star-forming and quiescent galaxies. However, there is some contamination from the star-forming population into the high \textit{NUV}-\textit{r}  tail and vice versa \cite{Kaviraj2007_nuv}. To avoid potential contamination from dusty and ``green-valley" galaxies, we identified galaxies as star-forming with \textit{NUV}-\textit{r}  $<$ 3.5 and quiescent with \textit{NUV}-\textit{r}  $>$ 5.0. This resulted in 492 quiescent and 759 star-forming galaxies. 

Finally, we selected \textit{HST} observations that have imaging in the F814W filter, to ensure a consistent comparison to the HST-SPOGs sample. The number of galaxies with F438W, F814W and F160W observations was too small to create a robust 3-color comparison sample. For the F814W observations, we required that the exposure time is sufficient to obtain equal or better depth to the HST-SPOG snapshot observations. After trying different exposure time cuts, we limited the exposure time to 500s, 1500s and 1500s for ACS, WFC3 and WFPC2 observations respectively.

After this cut, we obtained 186 quiescent and 220 star-forming galaxies with stellar masses, spectroscopic redshifts, deep \textit{HST} F814W and SDSS \textit{i}-band observations.  

\begin{figure}
    \centering
    \includegraphics[width=\linewidth]{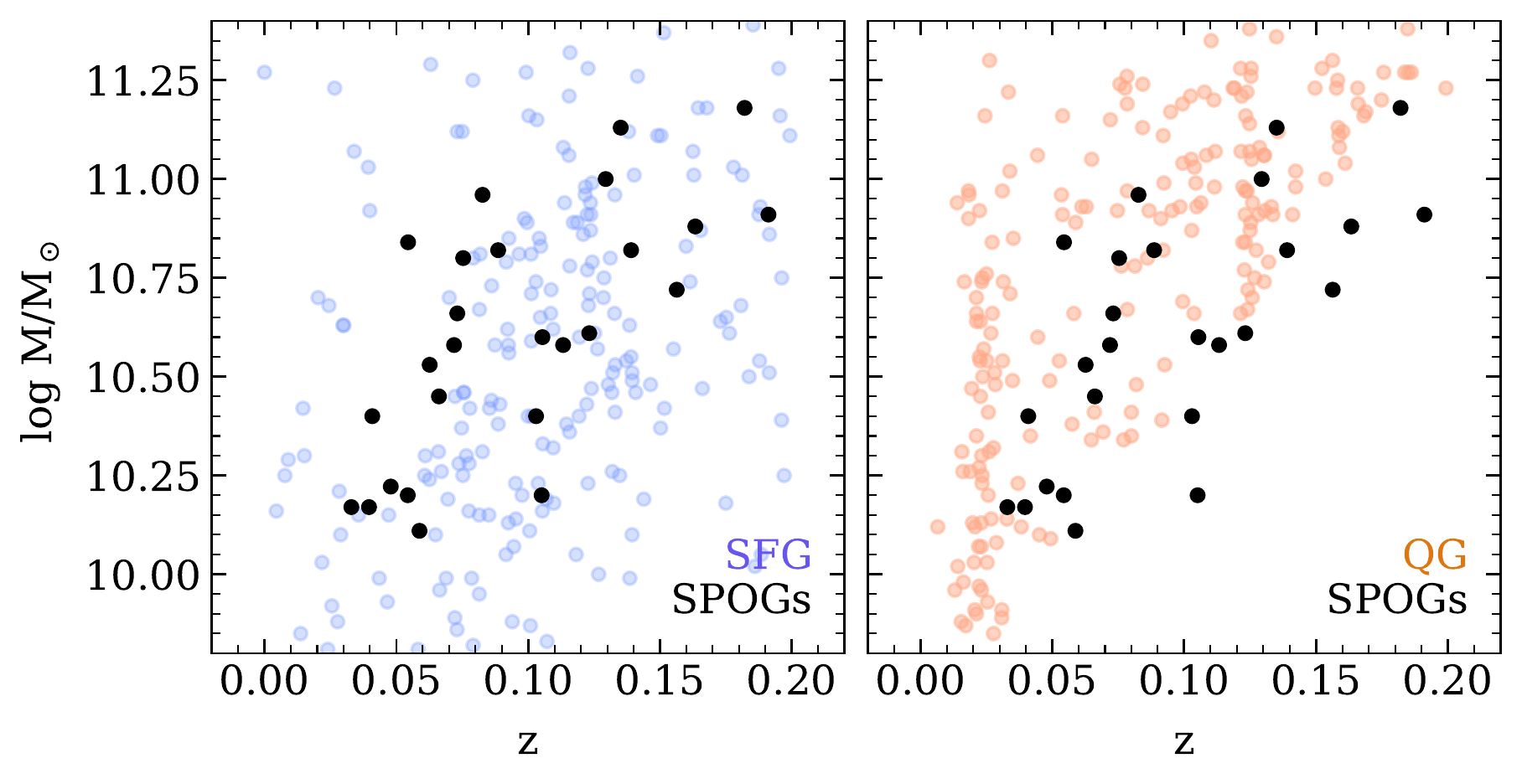}
    \caption{Mass and redshift distribution of the candidate comparison galaxies and HST-SPOGs (black). \textbf{Left:} candidate star-forming galaxies (blue), \textbf{Right:} candidate quiescent galaxies (orange).}
    \label{fig:mass_z_control}
\end{figure}

We examined each cutout visually. Since the selected galaxies were often simply foreground or background objects and not the primary targets of the \textit{HST} program, some galaxies were located on the image edges and therefore did not have reliable imaging data. Moreover, some galaxies were contaminated by a foreground object and some galaxies in the star-forming sample were ongoing mergers.

As mentioned above, the purpose of these comparison samples is to calibrate morphology metrics of HST-SPOGs against regular galaxies. Therefore, we removed any significantly disturbed galaxies from the calibration samples (i.e., those with clear companion, foreground or background contamination, or significant ``train-wreck" tidal features). We note that this way, our comparison sample is different from the general galaxy population, as in general $\sim$3\% of low-redshift galaxies are expected to be major mergers \citep[visual classification;][]{Darg2010}. This construction of the comparison sample enables us to predict the \textit{absolute} fraction of disturbed galaxies in the HST-SPOGs sample, rather than that relative to the general population. While we cannot tell whether mergers are \textit{more common} in HST-SPOGs than in the field, we can estimate the \textit{absolute} fraction of disturbed galaxies in HST-SPOGs, and use this to compare to merger fractions from other studies \citep[e.g.,][]{Ellison2008,Darg2010,Patton2011,Willett2013,Casteels2014,Man2016}. We removed 5 galaxies that had an extremely nearby companion making distinguishing them challenging, 124 galaxies that were either on the edge of the image or had insufficient depth in the image region the galaxy was located in, and 16 clear mergers, which agrees well with the $\sim$3\% major merger fraction in the field.

The remaining 138 quiescent and 157 star-forming galaxies formed a pool of our comparison sample candidates. The stellar mass and redshift distributions of the comparison sample candidates and HST-SPOGs are shown in Fig. \ref{fig:mass_z_control}. Note the lack of low-mass, higher-redshift quiescent galaxies, which are more difficult to detect and hence rarer. We then selected the comparison sample by finding one quiescent and one star-forming galaxy that optimally matched each HST-SPOG in terms of stellar mass and redshift.


In general it is possible that one comparison galaxy is the best match for more than one HST-SPOG. A repeated comparison galaxy is particularly likely in regions where no good matches are available, such as in the low mass quiescent regime. It is preferable to avoid using one comparison galaxy for more than one HST-SPOG to avoid correlations in our analysis. Therefore, instead of simply matching each HST-SPOG to its closest comparison candidate, we used an algorithm to minimize the \textit{total} difference in mass and redshift across all matches.

This problem is solved by the ``Hungarian algorithm" \citep{Kuhn1955}. Historically, this algorithm was developed to assign $N$ tasks to $N$ choices of people in a way to optimize the total cost to perform the tasks. The Python implementation of the Hungarian algorithm in the SciPy library allows doing this for unequal number of tasks $N$ and choices $M$, as long as $N > M$. This is done using an $N \times M$ cost matrix $D$, where  $D_{ij}$ corresponds to performing the $i$\ts{th} task by the $j$\ts{th} person.

In this case, we needed to assign $N$ available comparison candidates (quiescent or star-forming) to $M$ HST-SPOGs, where $N > M$. We did this as to minimize the total distance (i.e., cost matrix) in the mass and redshift space. We calculated $D_{ij}$ as the normalized distance in redshift and logarithmic mass space:
\begin{figure}
    \centering
    \includegraphics[width=\linewidth]{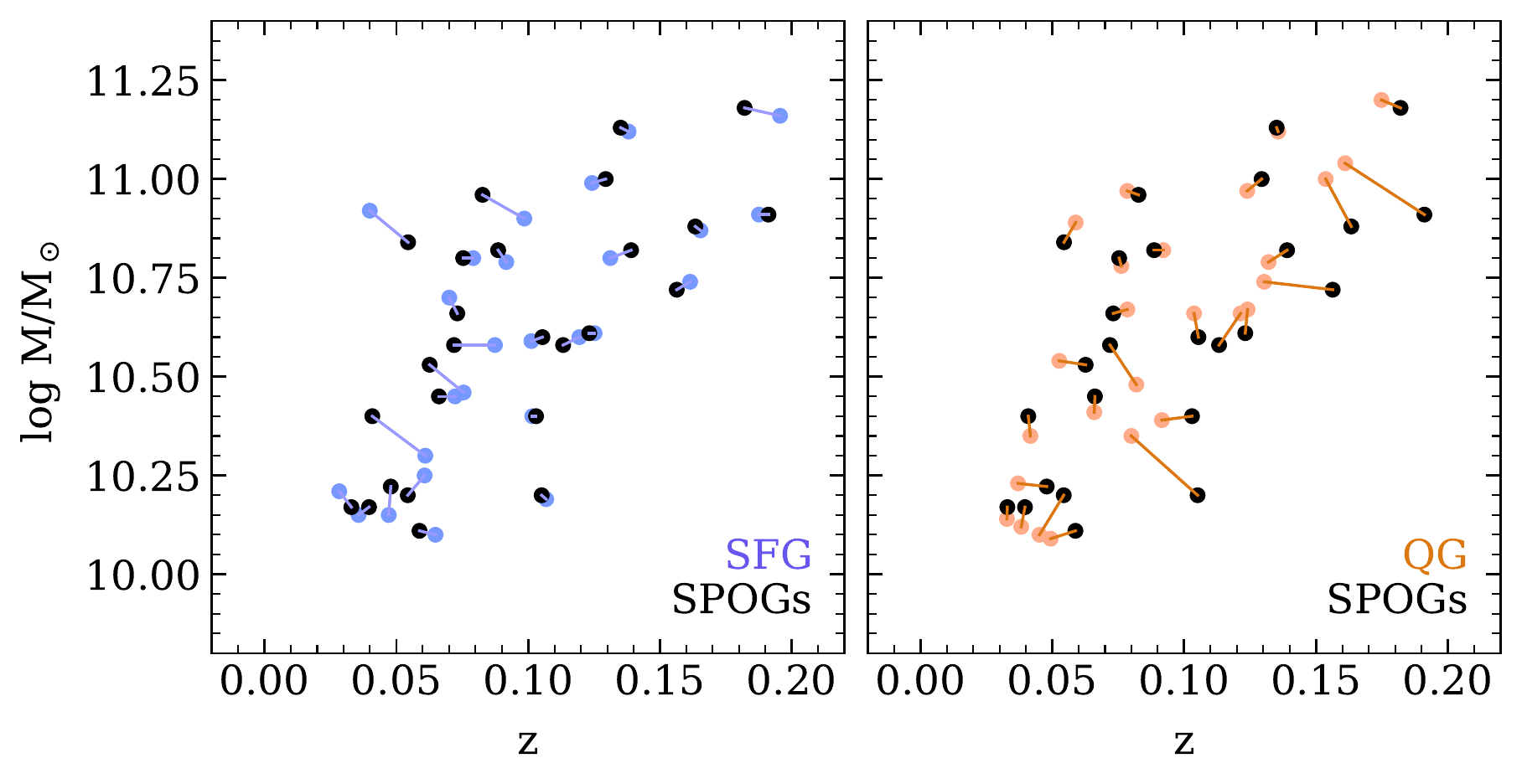}
    \caption{Mass and redshift distribution of the selected comparison galaxies and HST-SPOGs (black). \textbf{Left:} comparison star-forming galaxies (blue), \textbf{Right:} comparison quiescent galaxies (orange). Solid lines connect a comparison galaxy to its corresponding HST-SPOG.}
    \label{fig:mass_z_control_matched}
\end{figure}


\begin{figure*}
    \centering
    \includegraphics[width=\linewidth]{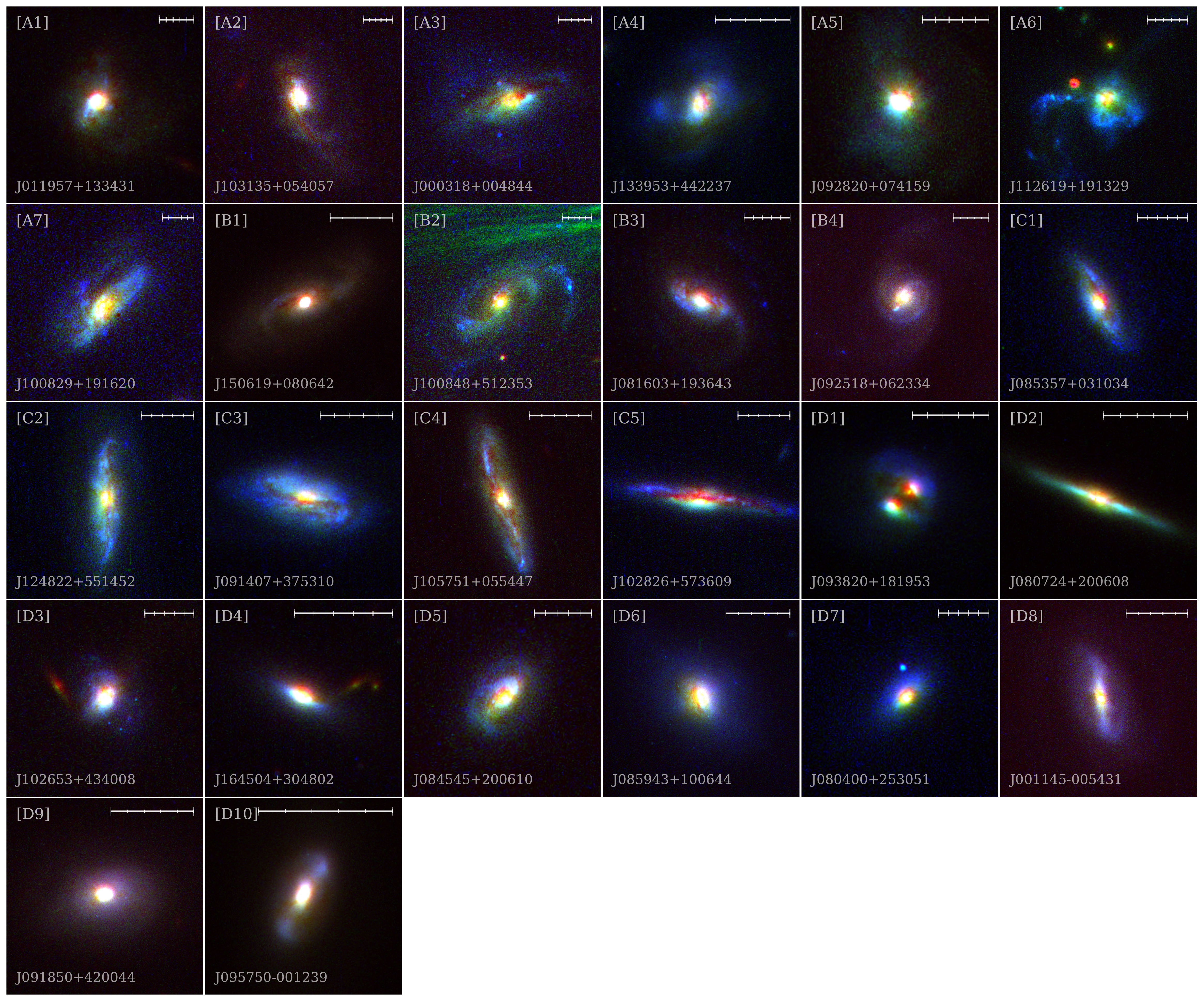}
    \caption{False-color images of 26 HST-SPOGs constructed using \textit{HST} F160W, F814W, and F438W observations as red, blue, and green channels respectively. The cutout size is 1.5$R_{p}$ for each galaxy. The white bar shows a 5 kpc scale. The galaxies are grouped visually into 4 groups as described in Sec. \ref{sec:visual}: mergers (A), merger candidates/S(B)c spirals (B), edge-on dusty spirals (C) and other (D). F438W and F814W imaging of galaxy B2 is contaminated by an artifact.}
    \label{fig:thumbnails}
\end{figure*}

\begin{equation}
    D_{ij} = \sqrt{\bigg( \frac{\log M^{c}_i - \log M^{s}_j}{\sigma_{\log M,s}} \bigg)^2 + \bigg( \frac{z^c_i - z^s_j}{\sigma_{z,s}} \bigg)^2},
\end{equation}
where for a pair of a comparison galaxy $i$ and an HST-SPOG $j$, $\log M_i^c$ and $\log M_j^s$ are the stellar mass logarithms, and $z_i^c$ and $z_j^s$ are the redshifts respectively. $\sigma_{\log M,s}$ and $\sigma_{z, s}$ are the standard deviation of the stellar masses and redshifts in the HST-SPOGs sample used for normalization.

We assigned the comparison sample galaxies to each HST-SPOG as to minimize the total cost function. The resulting samples of comparison quiescent and star-forming galaxies, with pair-wise assignment to the HST-SPOGs sample, are shown in Fig. \ref{fig:mass_z_control_matched}. 

As mentioned above, finding perfect matches for high-redshift, low-mass SPOGs was challenging, especially among the quiescent population. Our comparison QG sample generally has slightly larger masses and lower redshifts than the HST-SPOG sample. However, the difference is not significant: all of HST-SPOG, SFG, and QG samples have average mass $\log M_\star/M_\odot = 10.6^{+0.4}_{-0.3}$ and redshift $z=0.08^{+0.05}_{-0.03}$ where the uncertainties represent the 16\ts{th}/84\ts{th} percentiles of the median of each distribution after bootstrapping the distributions 10,000 times.

\section{Data Preparation}\label{sec:data_prep}

The SDSS observations were obtained from the SDSS Science Archive Server\footnote{\href{https://dr16.sdss.org}{SDSS DR16 Science Archive Server -- dr16.sdss.org}}, and \textit{HST} observations were obtained from the Mikulski Archive for Space Telescopes (MAST)\footnote{\href{https://archive.stsci.edu/access-mast-data}{MAST data access -- archive.stsci.edu/access-mast-data}}. Both SDSS and \textit{HST} images were reduced by the standard pipelines before being uploaded to the archive, as described on the Hubble Legacy Archive\footnote{\href{https://hla.stsci.edu}{Hubble Legacy Archive -- hla.stsci.edu}} for \textit{HST}, and in \cite{Stoughton2002} for SDSS. 

For both the HST-SPOGs and the comparison samples, we created cutouts of \textit{HST} and SDSS imaging centered on each galaxy. We used the galaxy Petrosian radius $R_{p}$ \citep{Petrosian1976} obtained from the SDSS photometric catalog and created a cutout with a size of $3R_p$. For some galaxies, the Petrosian radius estimate was incorrect, in which case we manually input a radius that produced a good cutout. 

\subsection{Snapshot \textit{HST} Imaging}

The \textit{HST} observations of 26 CO-detected SPOGs were obtained in a snapshot program (Proposal 14649; PI: Alatalo). For each galaxy, a single exposure was taken with a F438W, F814W, and F160W filters. The average image depth for each filter are given in Tab. \ref{tab:data_summary}.

Since only single exposures were taken, the standard pipeline reduction did not remove cosmic rays in F438W and F814W exposures, so cosmic rays were subsequently detected and interpolated over using the Python implementation of the LACosmic algorithm\footnote{\href{https://github.com/astropy/astroscrappy}{Astro-SCRAPPY -- github.com/astropy/astroscrappy}} \citep{vanDokkum2001}. Since F160W exposures were taken in MULTIACCUM mode, subsequent frames were used in cosmic ray rejection for the F160W data, and additional processing was not required.

F814W and F438W imaging has a finer pixel scale of 0.039\arcsec compared to the 0.128\arcsec pixel scale of F160W observations. To generate false-color images and to compute galaxy \textit{B}-\textit{H} and \textit{I}-\textit{H} colors, we re-projected the F160W imaging onto the finer grid using the \texttt{reproject} package. We chose to re-project \textit{H-}band data onto the finer grid of \textit{I}- and \textit{B}-band data because we did not perform any quantitative analysis on the color images that would require minimizing the interpolation error, and the finer grid allowed us to highlight small features in the false-color imaging.

Fig. \ref{fig:thumbnails} shows the false-color thumbnails of the HST-SPOG sample constructed with \textit{HST} F160W, F814W, and F438W imaging as red, green, and blue channels respectively. All images are scaled with an inverse hyperbolic sine scaling to highlight faint features, and an unsharp mask was applied for contrast. The galaxies are grouped visually as described in Section \ref{sec:visual}. The image of galaxy B2 is contaminated by an artifact in F814W and F438W imaging, masked in further analysis. 

\begin{figure}
    \centering
    \includegraphics[width=\linewidth]{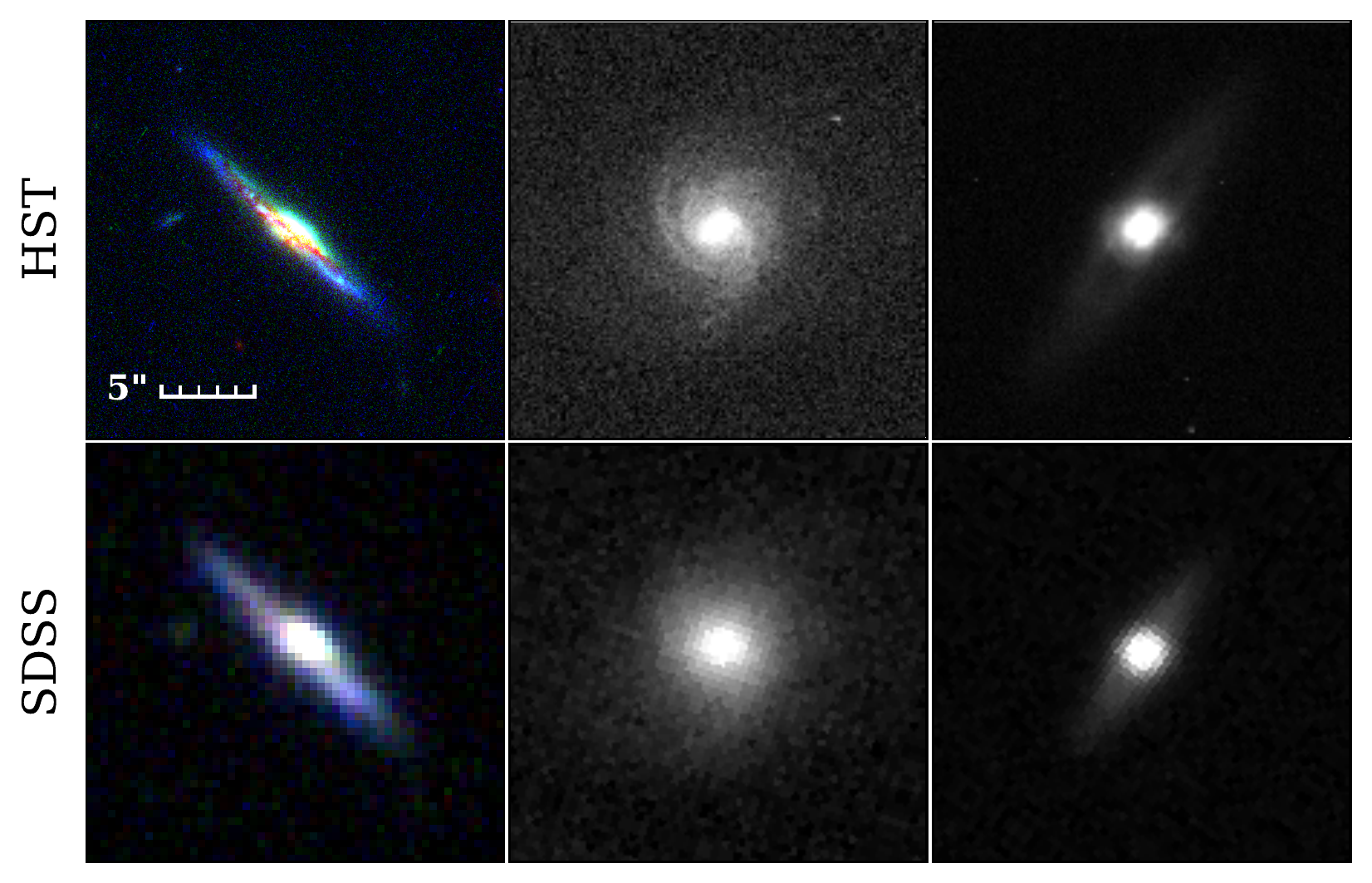}
    \caption{Thumbnails of an example HST-SPOG J102826+573609 (left), its matched star-forming (center) and quiescent (right) galaxies using \textit{HST} (top) and SDSS (bottom) imaging. All images are scaled to contain 1.5$R_p$ of the HST-SPOG. False-color images for the HST-SPOG are constructed with \textit{HST} F160W, F438W, F438W and  SDSS \textit{gri} imaging. Similar figures for the remaining HST-SPOGs are provided in the figure set in the online journal (26 images).}
    \label{fig:control_thumbnails}
\end{figure}

Fig. \ref{fig:control_thumbnails} shows \textit{HST} and SDSS thumbnails of an example HST-SPOG and its matched comparison star-forming and quiescent galaxies. The \textit{HST} image of the HST-SPOG is created the same way as in Fig. \ref{fig:thumbnails}. All three thumbnails are scaled to the same size, given by 1.5$R_{p}$ of the HST-SPOG, and rotated to align with the celestial north. The complete figure set showing similar plots for each HST-SPOG (26 images) is available in the online journal.

\subsection{\textit{HST} Background subtraction}
\label{sec:bg}

Archival \textit{HST} imaging contains a large number of photon counts due to the background sky emission. For the morphological analysis and consistency with SDSS imaging, we subtracted the mean sky background from the images. 

We estimated the mean background level using a sigma-clipping technique. First, the mean and the standard deviation $\sigma$ of all pixels is calculated. Then, pixels whose value exceeds $3\sigma$ are masked, and the standard deviation is computed again. This process is repeated until convergence, i.e. no additional pixels are masked. 

The mean background level $\mu_{bg}$ and its standard deviation $\sigma_{bg}$ are computed using the masked image. We then subtracted $\mu_{bg}$ from all image cutouts.

\subsection{Comparison sample and SPOG depth matching}
\begin{figure*}
    \centering
    \includegraphics[width=\linewidth]{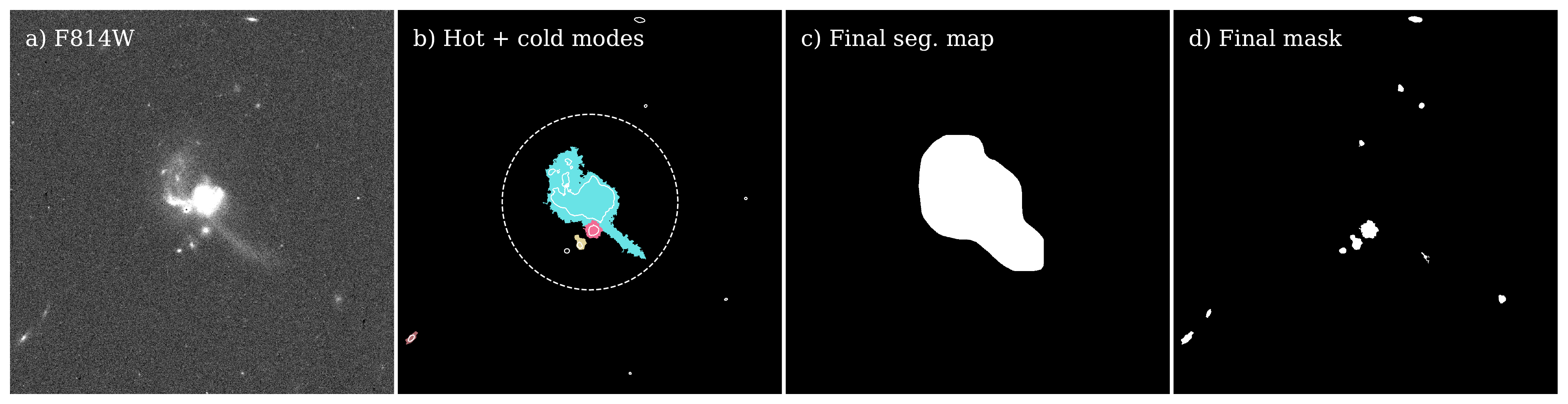}
        \caption{Example segmentation algorithm for F814W imaging of J112619+191329. \textbf{a)} original imaging; \textbf{b)} the ``cold" segmentation map with a random color scheme for distinct segments and ``hot" contours overplotted in white; \textbf{c)} the final galaxy segmentation map; \textbf{d)} the mask containing all other contaminant objects. The dashed line in panel (b) shows a $2R_{p}$ radius used to exclude faint outer contaminants. The aim of the segmentation process is to deblend all potential contaminants from the galaxy envelope while keeping tidal features undeblended. Since it is challenging to determine whether a bright object is a contaminant or physically associated with the galaxy, we strictly remove all potential contaminants for purity. The detailed description of the algorithm, and the implications for the morphological measurements, are given in Sec. \ref{sec:segmentation}.}
    \label{fig:segmap}
\end{figure*}
As described in Section \ref{sec:control}, we selected comparison galaxies that are as deep or deeper than HST-SPOGs \textit{HST} imaging. To ensure a robust comparison, we then matched the depth of the comparison galaxy cutouts to the depth of HST-SPOGs cutouts. The image limiting depth $M_\sigma$ in magnitudes is then given as
\begin{equation}
    M_\sigma = -2.5 \log{\sigma_{bg}} + \textrm{ ZP},
\end{equation}
where ZP is the photometric zeropoint of the image. Since the cutout size was $3R_{p}$, enough of the image did not contain any galaxy light above the noise limit to allow this calculation. We computed the average depth of HST-SPOGs F814W imaging from individual $\sigma_{bg}$ estimates to be $M_\sigma^s = 22.39 \pm 0.06$ mag/arcsec\ts{2}, so we chose to match all comparison \textit{HST} observations to this depth.

Some comparison galaxies were located on the edges of the \textit{HST} images, where the drizzled archival exposures were not aligned, so the effective image depth in those regions was shallower than the overall observation. We used the context information provided by the \textit{HST} reduction pipeline to determine the region that has the same total exposure time as the comparison galaxy. We then only calculated the background noise in that region.

To match the image depth to the average depth of SPOG images, we then added extra white noise using a Gaussian white noise distribution with a 0 mean and a standard deviation $\sigma'$ given by
\begin{equation}
    \sigma'^2 = \sigma^2 \Bigg( 10^{\frac{2(M_\sigma^c - M_\sigma^s)}{2.5}} - 1 \Bigg),
\end{equation}
where $M_\sigma^c$ is the comparison image depth and $\sigma$ is the background standard deviation of the original comparison image.

\subsection{Point Spread Function}\label{sec:psf}

We used the point spread function (PSF) of each image to ensure a robust morphological analysis and account for seeing effects. For \textit{HST} imaging, we constructed a simulated PSF using the TinyTim software \citep{Krist2011}. For SDSS imaging, we used the empirical PSF measurements that are publicly available for each SDSS field on the SDSS Science Archive Server.

\subsection{Image segmentation}\label{sec:segmentation}

To analyze the galaxy morphology, we first needed to create segmentation maps showing which pixels belong to the galaxy object and which are background pixels or foreground/background objects. To do this, we adapted the segmentation routine from the \texttt{photutils} package \citep{photutils}. The implementation of \texttt{photutils} closely matches that of the \textsc{SExtractor} software \citep{Bertin1996} commonly used in similar applications. 

A correct segmentation map identifying the galaxy is crucial to calculate robust morphological measurements. The aim is to exclude as many foreground and background objects as possible while including tidal or other disturbed features in the same galaxy map. This is challenging in practice, as separating overlapping distinct objects requires a high degree of deblending, which can also  deblend the galaxy and its tidal features. We have specifically tuned our segmentation algorithm to maintain as many tidal features and removing foreground/background objects as strictly as possible. As a result, if a galaxy has merging companions or large star clusters, they will very likely be deblended and detected as a separate object. However, since we cannot tell whether an object is a companion or a foreground/background contaminant, we chose to remove all potential contaminants. This decreases the power of some morphological metrics that are specifically tuned to look for doubly nucleated galaxies or enveloped companions \citep[e.g., Gini and M$_{20}$,][]{Lotz2004} but does not affect the metrics that look for wide-spread disturbances, such as asymmetry.

The segmentation process can be adjusted using three main \texttt{photutils} input parameters: \texttt{NPIXELS}, the minimum region size in pixels; \texttt{NSIGMA}, the detection threshold in standard deviations, and \texttt{CONTRAST}, the flux contrast required to deblend two overlapping sources.  \cite{Galametz2013} created galaxy segmentation maps with \textsc{SExtractor} by iterating the segmentation process twice, creating a ``hot" segmentation map, identifying brightest regions, a  ``cold" map, identifying fainter features, and then combining the two. Similarly, we iterated the segmentation process three times in three different modes. Each time, we used a 2D tophat smoothing kernel with a 5 pixel kernel size to smooth over the pixel-to-pixel noise. The input parameters used for each mode are given in Tab. \ref{tab:segmap}.

\begin{deluxetable}{rccc}
\label{tab:segmap}
\tablecaption{\texttt{photutils} parameters used in producing the galaxy segmentation maps.}
\tablehead{
\colhead{Parameter} & \colhead{``Hot" mode} & \colhead{``Cool" mode} & 
\colhead{``Cold" mode}
}
\startdata
\texttt{NSIGMA} & 97\ts{th} percentile & 1$\sigma$ & 1$\sigma$ \\
\texttt{NPIXELS} & 1 px & 1 sq. arcsec & $0.01 R_{p}^2$ \\
\texttt{CONTRAST} & \nodata & $10^{-6}$ & \nodata \\
\texttt{NLEVELS} & \nodata & 32 & \nodata \\
\texttt{MASK} & Original & Original & Object mask \\
\texttt{FILTER\_KERNEL} & \multicolumn{3}{c}{Tophat kernel with $R=5$ pixels}
\enddata
\end{deluxetable}

\textbf{Step 1: ``hot" mode.} First, we created a ``hot" map to identify all bright objects in the image, using a 1 pixel minimum area, a large threshold $S/N$ and no deblending. The purpose of the ``hot" mode is to detect bright contaminating sources in the image. Since the $S/N$ threshold is high, separate objects are deblended by construction since we mask out the diffuse light that they may be enveloped in. This is shown as the white contours in Fig. \ref{fig:segmap}b. 

\textbf{Step 2: ``cool" mode.} Second, we created a ``cool" map with a $1\sigma$ threshold, 1 sq. arcsec minimum area, and a $10^{-6}$ deblending contrast.  This selected extended objects in the image, each shown as differently colored regions in Fig. \ref{fig:segmap}b. Note that the white contours are located inside the extended colored regions, indicating brightness peaks of the extended objects. We kept the minimum area large to avoid deblending the galaxy envelope, so not all faint background objects are detected in this pass. 

\textbf{Step 3. Mask contaminants.} We then used the ``hot" and ``cool" maps to mask objects that do not belong to the central galaxy. First, we found the segment coinciding with the galaxy center and labelled it as the ``main" segment, which is not masked. Then, we masked all other ``cool" segments that were sufficiently far away from the galaxy center, using threshold $R_{\textrm{thres}} = 2R_{p}$ shown as the dashed line in Fig. \ref{fig:segmap}b. Finally, we needed to mask any contaminants that are near or embedded in the galaxy. We found all ``hot" regions (other than the galaxy center) embedded in each ``cool" region. Then, if the ``cool" region covers more than 80\% of any ``hot" region, we masked it, since it contains a bright contaminant peak. This way, we masked all contaminant objects.

\textbf{Step 4: ``cold" mode.} Finally, we ran a third segmentation algorithm, using the 0.01$R_{p}^2$ as an area threshold, 1$\sigma$ brightness threshold, the mask from step 2, and no deblending. Since we already masked contaminants near the center of the galaxy, deblending was not needed, as we can assume that all of the unmasked light belongs to the galaxy itself. We used a lower area threshold than in step 2 to finally detect all faint and small objects away from the galaxy center and mask them. The resulting segmentation map consisted of a large region corresponding to the galaxy of interest, and small faint objects that have been missed by the ``cool" mode earlier. We masked all regions except the central galaxy. Finally, we smoothed the galaxy segmentation map using a uniform filter equal to 10\% of the image size to regularize it. This resulted in the final segmentation map, shown in Fig. \ref{fig:segmap}c, and the corresponding final mask, shown in Fig. \ref{fig:segmap}d. 

Note that the final mask may overlap the final segmentation map (e.g., if a foreground star is embedded in the galaxy halo). Both the mask and the segmentation map are passed to any further analysis tool, so all masked regions are not considered in morphological analysis, even if embedded in the segmentation map.

\section{Morphology Measurements}\label{sec:morphology}

The morphological analysis of HST-SPOGs was carried out in three different ways: 1) visual analysis and description of the HST-SPOG snapshots in Sec. \ref{sec:visual}, 2) non-parametric analysis with a computational morphology code \textsc{statmorph}, and 3) parametric analysis using a S\'ersic profile fitting tool \textsc{GALFIT}.  Steps 2 and 3 were performed with \textit{HST} F438W, F814W and F160W imaging and SDSS \textit{i}-band imaging of the HST-SPOG sample, as well as \textit{HST} F814W and SDSS \textit{i}-band imaging of the comparison samples.

Both \textsc{statmorph} and \textsc{GALFIT} allow the user to compute a large variety of morphological measurements. Detailed descriptions of all of the various parameters of \textsc{GALFIT} and \textsc{statmorph} can be found in \cite{galfit} and \cite{statmorph} respectively. In the sections below, we define only the parameters used in this work.

\subsection{Non-parametric morphology with \textsc{statmorph}}

\textsc{statmorph} is an open-source non-parametric morphology calculation library. For every galaxy image, it computes a variety of morphology measurements, including S\'ersic index and radius \citep{Sersic1963}, concentration, asymmetry, smoothness \citep{Conselice2003}; Gini and M$_{20}$ \citep{Lotz2004}; multimode, intensity and deviation \citep{Freeman2013}; and shape asymmetry \citep{Pawlik2016}. We note that there are subtleties in the computational methods of all morphological measurements, so it is crucial to be careful comparing morphology values across studies that use different morphological software.

To run \textsc{statmorph}, we use the science image of the galaxy, its segmentation map (Sec. \ref{sec:segmentation}), and a mask identifying  extraneous objects, bad pixels and cosmic rays. In addition, \textsc{statmorph} convolves the model image with a PSF to improve the accuracy of the S\'ersic fit. We supply the algorithm with the PSF computed in Sec. \ref{sec:psf}. Below are the parameters we used in this work that were obtained with \textsc{statmorph}.

\subsubsection{Concentration}

The concentration of a galaxy's light distribution can be measured non-parametrically, without assuming a model for the distribution like a S\'ersic profile. The most common way to measure concentration has been proposed by \cite{Bershady2000}, where concentration ($C$) is defined as the ratio between the galaxy's 80\% and 20\% circular isophotes:

\begin{equation}
    C = 5 \log_{10} \frac{r_{80}}{r_{20}},
\end{equation}
where $r_{80}$ and $r_{20}$ contain 80\% and 20\% of the galaxy's total flux, respectively. High values of concentration imply that these radii are nearby, and therefore most of the flux is constrained to the center. Galaxies with $C \approx 5$ have bulge-dominated morphologies and S\'ersic indices $n \approx 4$.

However, since this measurement uses circular isophotes, it assumes a circular symmetry. It does not account for light distributions that are strongly edge-on or disturbed.

\subsubsection{Asymmetry}

The asymmetry of the galaxy's light distribution can identify galaxies that are in ongoing mergers or that have post-merger signatures such as tidal tails and double nuclei. Asymmetry ($A$) is measured by rotating the galaxy image by 180$^{\circ}$ and subtracting the original and the rotated distributions \citep{Schade1995, Abraham1996, Conselice2003}:

\begin{equation}
    A = \frac{\sum_{ij} |F_{ij} - F_{ij}^{180}|}{\sum_{ij} |F|} - A_{bg},
\end{equation}
where $F_{ij}$ is the flux in a given pixel, $F_{ij}^{180}$ is the flux of the rotated image in the same pixel location, and $A_{bg}$ is the asymmetry of the background. The center of rotation is chosen iteratively to minimize the $A$. Galaxies with $A > 0.1$ are generally highly asymmetric, although this depends strongly on image quality used in calculating asymmetry \citep{Lotz2004}. \textsc{statmorph} calculates asymmetry within 1.5$R_p$, which may differ from other software implementations \citep[e.g., $R_{max}$ in][]{Pawlik2016}. The size of the aperture determines the contribution of $A_{bg}$ term, and therefore asymmetry values computed within larger radii will be smaller, even if other image parameters are the same.

\subsubsection{Shape Asymmetry}

Shape asymmetry $A_S$ is computed the same way as asymmetry, except using the binary detection map identifying pixels belonging to a galaxy rather than the image itself \citep{Pawlik2016}. The shape asymmetry segmentation map aims to specifically detect faint and irregular features in the outer region on the galaxy, and is not smoothed as much as the segmentation map described in Sec. \ref{sec:segmentation}. The detection map is calculated by \textsc{Statmorph} and differs from the user-input segmentation map. The purpose of the shape asymmetry detection map is to pick out all faint and irregular features in the outer regions of the galaxy, therefore they are created with less smoothing than segmentation maps described in Sec. \ref{sec:segmentation}.

Since $A_S$ weights all galaxy pixels the same way, it is more sensitive to fainter tidal features rather than asymmetry in the galaxy center and can be a better metric to identify fading post-merger signatures. However, it is also extremely sensitive to the way the detection map is computed, and can be strongly affected by improper masking of contaminant objects or masking of galaxy's tidal features during segmentation.

\subsubsection{Half-light Radius}

We used an elliptical half-light radius, $R_{0.5}$. First, the center of the galaxy is found so as to minimize its asymmetry; and then the semi-major axis of an ellipse containing 50\% of the flux is used as $R_{0.5}$. 

\subsubsection{G-M$_{20}$ bulge strength \& disturbance}

The Gini index ($G$) and M$_{20}$ are both non-parametric measures of the concentration of a galaxy's light. $G$ measures how uneven the light distribution is without considering the location of the brightest pixels. Therefore, it is not sensitive to offset bright regions unlike similar measurements (concentration or Sérsic index). On the other hand, M$_{20}$ measures the distance of the brightest 20\% of the light from the galaxy center and therefore is especially sensitive to non-central bright regions.

When used in tandem, they can evaluate the degree of bulge strength and disturbance of a galaxy \citep{Lotz2004, Snyder2015b, statmorph, Sazonova2020}. Regular late- or early-type galaxies lie along the ``G-M$_{20}$ main sequence". \cite{statmorph} define this sequence for galaxies with $\log_{10} M_\star / M_\odot \approx 9.8 - 11.3$ and $z \approx 0.05$, which fits well the HST-SPOGs sample. The main sequence is defined as:
\begin{equation}
    F(G, M_{20}) = -0.693 M_{20} + 4.95G - 3.96
\end{equation}

Bulge-dominated galaxies have centrally concentrated light with a low $G$ and low M$_{20}$, leading to a high $F$;  while disks have diffuse light with bright offset knots of star formation, leading to a high $G$, high M$_{20}$, and low $F$. To avoid unnecessary abbreviations, we refer to $F(G, M_{20})$ as ``G-M$_{20}$ bulge strength" for the remainder of this paper.

Deviations from the ``main sequence" can be indicative of disturbed features. For example, galaxies with a low $G$ but a high M$_{20}$ have highly concentrated bright regions that are off-center. This is typical of late-stage mergers with double nuclei \citep{Snyder2015a}. This disturbance is measured as the offset from the main sequence:

\begin{equation}
    S(G, M_{20}) = 0.139M_{20} + 0.99G - 0.327;
\end{equation}
we refer to $S(G, M_{20})$ as ``G-M$_{20}$ disturbance" for the remainder of this paper.

\subsection{Sérsic morphology with \textsc{GALFIT}}

\subsubsection{Sérsic Index}

Se\'rsic radius and index \citep{Sersic1963} are commonly used parametric measures of galaxy morphology. The galaxy's light distribution can be approximately modelled by a S\'ersic profile:
\begin{equation}
    I(R) = I_{0.5} \exp \Bigg\{ -b \Bigg[ \bigg( \frac{R}{R_{S,0.5}} \bigg)^{1/n} - 1\Bigg] \Bigg\},
\end{equation}
where $R_{S,0.5}$ is the effective radius containing half of the total light, called the S\'ersic radius;  $I_{0.5}$ is the intensity at $R_{S,0.5}$; $n$ is the S\'ersic index; and $b$ is a function of $n$ computed via Gamma functions. In the local Universe, elliptical galaxies are well-described by S\'ersic indices $n = 4$ \citep{deVaucouleurs1948}, while spiral galaxies have $n = 1$ or an exponential profile \citep{Freeman1970}. Previous studies show that, on average, PSBs have intermediate S\'ersic indices, indicating an ongoing transition from a disk-like to a spheroidal morphology \citep{Blake2004, Yang2004, Yang2008, Pawlik2016, Pawlik2018, Chen2019}.

In addition to the Sérsic index and radius, Sérsic fitting produces estimates of the galactic center (center of the modelled light distribution), ellipticity, and orientation.

\subsubsection{\textsc{GALFIT}}\label{sec:galfit}

Although \textsc{statmorph} can perform one-component S\'ersic fits, \textsc{GALFIT} \citep{galfit} is a much more versatile tool to perform S\'ersic modelling. Unlike \textsc{statmorph}, \textsc{GALFIT} allows the user to perform multi-component fitting, as well as incorporating models other than the S\'ersic one and various Fourier modes to capture perturbations in the galaxy light distribution.

However, the complexity of the optimization problem rises significantly when more components are added to the fit, so precise modelling of galaxy light requires manual inspection and careful choice of the initial parameter guesses. Instead, to ensure a robust comparison across the galaxies in the SPOG and comparison samples, we opted for using simpler models and using the residuals to characterize the disturbance.

We performed two different fits. First, we used a Sérsic fit on all \textit{HST} and SDSS imaging. For the \textit{HST} imaging of HST-SPOGs and comparison galaxies, we included a central Gaussian component with a varying size $\lambda_{\textrm{PSF}} \leq R \leq 2\lambda_{\textrm{PSF}}$ to model central emission. This is necessary because HST provides a high enough spatial resolution for nearby galaxies that single- and two-component Sérsic fits are insufficient to model the complex galaxy substructures. In particular, many nearby galaxies require an additional compact central component to model the emission from the galaxy core \citep[e.g.,][]{galfit}. Excluding the central source leads to \textsc{Galfit} overestimating the Sérsic index due to the cuspy nuclear emission, and hence producing high residuals. We did not include this component in SDSS fits as it is unnecessary for lower-resolution imaging. We used the results from the Sérsic fits (including the central component for \textit{HST} imaging) to estimate the Sérsic index $n$ for all galaxies in further analysis.

We then performed a multiple-component fit with a Sérsic disk ($n=1$) and a bulge component ($n=4$) and used this to calculate the bulge-to-total light ratio, $B/T$. Again, we included a Gaussian PSF component in \textit{HST} imaging of HST-SPOGs and comparison galaxies. When calculating bulge-to-total light ratio, we added the Gaussian PSF flux to the bulge flux.

In general, galaxies consist of a disk and a bulge, so multi-component fits perform better on data with high spatial resolution \citep{galfit}. Since our sample has a low redshift, the \textit{HST} imaging has high enough resolution that two-component fits are better. SDSS imaging, on the other hand, has low resolution due to atmospheric seeing, so two-component models overfit the data and single-component models perform better. 

Finally, \textsc{GALFIT} reliability depends strongly on the initial guess passed by the user. We used the parameters from \textsc{statmorph} as the initial guess. During the multi-component GALFIT routine, we fixed the central location of the  components to be the same to reduce the degrees of freedom of the fit. 
As mentioned above, Sérsic fitting produces estimates of the galaxy center, half-light radius, ellipticity, and orientation. However, these values are difficult to calculate in multi-component models as they can differ for each model. Our sample contained very disturbed galaxies as seen in Fig. \ref{fig:thumbnails}, and even multi-component models did not always provide good fits, so these estimates were not always representative of the galaxies' real properties. In further analysis (e.g., in measuring the residual flux fraction described below), we opted to use the non-parametric measurements of the galaxy center, half-light radius $R_{0.5}$, ellipticity, and orientation.

\begin{figure}
    \centering
    \includegraphics[width=\linewidth]{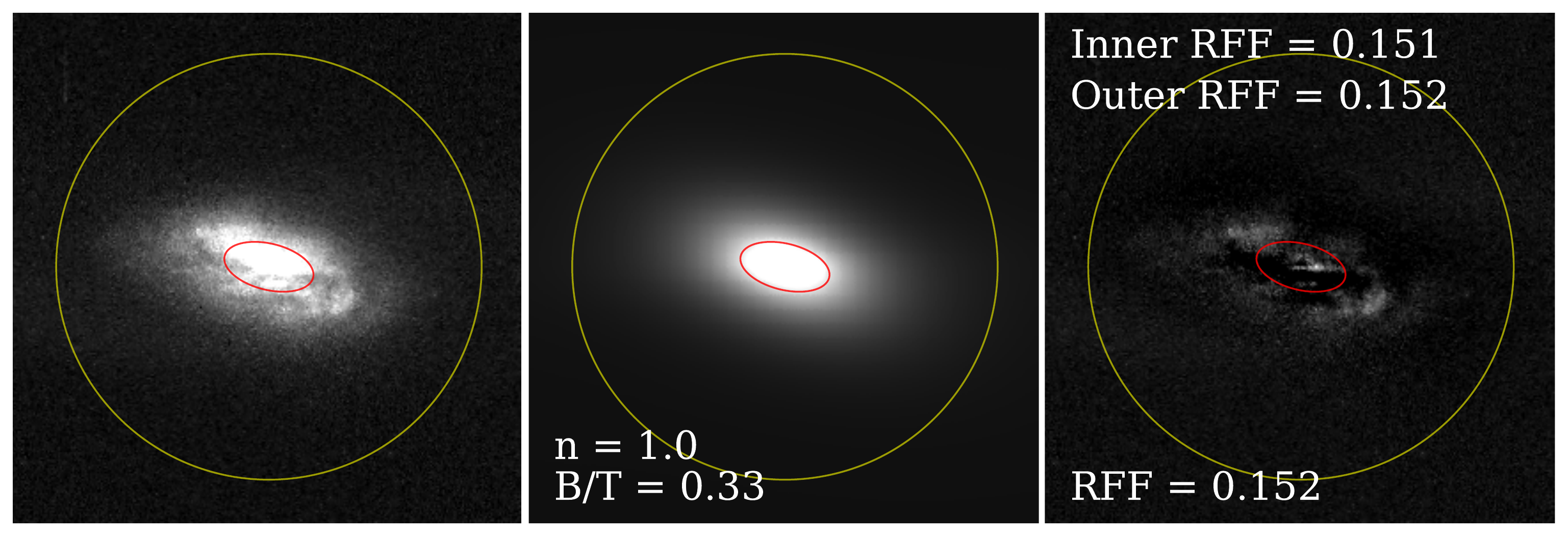}
    \caption{\textsc{Galfit} analysis of an example HST-SPOG J091407+375310. \textbf{Left:} \textit{HST} F814W image of the galaxy; \textbf{center:} \textsc{Galfit} disk + bulge + PSF model described in Sec. \ref{sec:galfit}; \textbf{right:} residual after the fit used to calculate the RFF. The yellow circle has radius $2R_{p}$ and shows the region where RFF is calculated. The red ellipse has the semi-major axis $R_{0.5}$ and shows the division between inner and outer RFF regions. Fit results for the remaining galaxies used in this paper are provided in the figure set in the online journal (208 images).}
    \label{fig:rff}
\end{figure}

\vspace{-1em}
\subsubsection{Bulge and disk decomposition}

Muti-component Sérsic fits can be performed to break the galaxy down into the bulge and the disk components. The bulge strength is then evaluated using the bulge to total light flux ratio. In our analysis, we performed a GALFIT fit using an $n=1$ exponential disk and a $n=4$ bulge ($B$) components. We then calculated the bulge-to-total ratio as:

\begin{equation}
    B/T = \frac{F_b}{F_t} = \frac{1}{1 + \big(F_b/F_d\big)^{-1}},
\end{equation}
where $F_b$ and $F_d$ are the model bulge and disk fluxes, respectively. $F_t$ is the total flux contained the model, not to be confused with the total galaxy flux. If a central PSF component is included in the fit, we set $F_b$ to the combined flux of the bulge and the PSF components. Higher values of $B/T$ correspond to more prominent bulge components, although some early-type S0 galaxies have weak bulges with $B/T$ near 0 \citep{FraserMcKelvie2018}. 

\subsubsection{Residual Flux Fraction}

The residual flux fraction (RFF) measures the fraction of the total galaxy's flux that is not well fit by a S\'ersic profile \citep{Hoyos2011, Hoyos2012}. It is calculated by performing the S\'ersic fit and then subtracting the model from the original image, resulting in a residual image. The RFF is then given by:

\begin{equation}
    \textrm{RFF} = \frac{\sum | F^{\textrm{res}} | - 0.8 \times \sum \sigma_{bg}}{\sum F^{\textrm{model}}},
\end{equation}
where $F^{\textrm{res}}$ is the residual flux, $\sigma_{bg}$ is the standard deviation of the image, and $F^{\textrm{model}}$ is the S\'ersic model flux. The RFF is calculated using the absolute of the residual image, since the residual can be both positive and negative depending on whether the model under- or over-fits the data. The background contribution to RFF is subtracted with the $0.8 \Sigma \sigma_{bg}$ term, as described below. 

We can break down the image flux at each pixel into the object flux, $f^{sci}$, and background flux, $f^{bg}$. Our aim is to subtract out the contribution of the background flux from the RFF measurement. Since the images are background-subtracted as described in Sec. \ref{sec:bg}, the average background flux is 0 and $f^{bg}$ follows a normal distribution: $f^{bg} \sim \mathcal{N}(0, \sigma_{bg})$. 

For calculations involving the sum of flux from all pixels (such as photometry), the total flux is simply the total object flux, since the expected total background flux is 
\begin{equation}
    F_{bg} = \sum f_{bg} = N \langle f_{bg} \rangle = 0,
\end{equation}
where $N$ is the number of pixels, and $\langle f_{bg} \rangle$ is the expectation value of the background flux, 0. 

However, for a sum of the \textit{absolute} flux at each pixel,
\begin{equation}
    |F_{bg}| = \sum |f_{bg}| = N \langle |f_{bg}| \rangle = 0.8 N \sigma_{bg},
\end{equation}
which can be shown by integrating the Gaussian noise distribution. Therefore, we must subtract $0.8N\sigma_{bg}$, the background contribution to the RFF.

In this analysis, we constrain the RFF measurement to $2R_{p}$ around the galaxy center. This is different from the Kron radius used by \cite{Hoyos2011, Hoyos2012}. We opted for the Petrosian radius, because it is robust to changes in redshift and is consistent with the radius used by non-parametric measurements from \textsc{statmorph}.

In addition, to differentiate the internal disturbance and tidal features, we calculated inner RFF\sub{in} and outer RFF\sub{out}. RFF\sub{in} is defined as the RFF within $R_{0.5}$, and RFF\sub{out} is defined as the RFF between $R_{0.5}$ and $2R_{p}$. We used the \textsc{Galfit} fit with the lowest $\chi^2$ (i.e., the best fit) to calculate the RFF for each image. An example of the \textsc{Galfit} two-component fit, the residual, and the RFF calculation is shown in Fig. \ref{fig:rff}. The figure set in the online journal shows the residuals for the remaining HST-SPOGs and comparison galaxies (208 images).

\subsection{Color gradients}\label{sec:color}

Galaxy color gradients can be a powerful probe into the mechanism behind galaxy quenching and the direction it acts in (inside-out or outside-in), as well as into the distribution of the galaxy dust content that traces the molecular gas.

We measured \textit{B}-\textit{I}, \textit{I}-\textit{H}, and \textit{B}-\textit{H} color gradients of HST-SPOGS using \textit{HST} F438W, F814W and F160W imaging. Since HST-SPOGs have different physical sizes, we computed color gradients as a function of $R_{0.5}$ instead of the galaxy physical size. We measured the average color in increments of $0.1R_{0.5}$ for $r\leq 2.5R_{0.5}$. At each radius $r$, we defined an elliptical annulus aperture around the galaxy center with bounding radii $r$ and $r+0.1R_{0.5}$, and the ellipticity and orientation given by the \textsc{Statmorph} $I$-band fit. We computed the weighted median, 16\ts{th}, and 84\ts{th} quantiles of the galaxy color in each aperture. We used flux $S/N$ ratio as a weight at each pixel. We then found the average color gradient of all HST-SPOG galaxies, normalized by $R_{0.5}$ of each galaxy.





\section{Results: Properties of HST-SPOGs}
\subsection{Visual properties \& groups}\label{sec:visual}

We used \textit{HST} imaging from Fig. \ref{fig:thumbnails} to visually classify the HST-SPOG sample in 4 groups. Group A (7 galaxies) represents clear post-merger galaxies with disturbed morphologies. Group B (4 galaxies) are post-merger candidates identified by possible tidal arms; but they could also be Hubble type S(B)c galaxies with loose spiral arms. Group C (5 galaxies) contains disk-like galaxies with prominent dust lanes embedded in the disk. Finally, group D (10 galaxies) contains all ``other" galaxies: they are mostly bulge-dominated, with dust-obscured nuclei but without clear dust lanes.

In Group A alone, 27\% (7/26) of our sample are clear post-mergers. This fraction is significantly higher than the 1.5--5\% typical for nearby field galaxies \citep{Darg2010} but lower than $\sim $50\% found for post-starbursts in previous studies \citep[e.g.,][]{Pawlik2016,Chen2019}, including the CO-SPOG parent sample \citep{spogs2}. Moreover, although Group A was selected as clear mergers, some of Group B galaxies are likely mergers (e.g., the features on B1 are likely tidal, and B4 has clear shells). Galaxies in Group D also show disturbed morphologies (D1, D3, D5, D8, D10). Under less conservative selection criteria, our sample contains at least 50\% of post-merger candidates, consistent with $46\pm7$ \% of the parent sample \citep{spogs2}. However, since the techniques used to identify disturbed galaxies in mergers across studies vary, comparing them directly is difficult, as discussed further in Sec. \ref{sec:disturbances}.

\textit{HST} imaging reveals that the galaxies in this sample are very dusty. HST-SPOGs are CO-selected and therefore have more molecular gas than the parent SPOGs sample. Depletion timescales for molecular gas and dust are correlated \citep{Michalowski2019,Li2019}, so CO-rich post-starbursts are expected to also have more dust. Galaxies in Group C have clear dust lanes typical of edge-on spiral galaxies. Galaxies in other groups are primarily face-on, so such dust lanes would be less apparent. However, all galaxies still show significant amounts of dust obscuration. Galaxies outside of Group C appear to have unusual dust morphologies that lead to a disturbed appearance of the galaxies — e.g., D1 and D10. Disrupted dust morphology can be linked to a disturbed molecular gas and possible outflows \citep[e.g.,][]{Obied2016, Kaufman2020}. 

Finally, the galactic nuclei of all HST-SPOGs are visibly redder than the outer regions. This could imply an older stellar population in the center and an inside-out quenching mechanism. However, considering the high amount of dust attenuation, the red colors are likely caused by dust obscuration in the nucleus. 

\begin{figure}
    \centering
    \includegraphics[width=\linewidth]{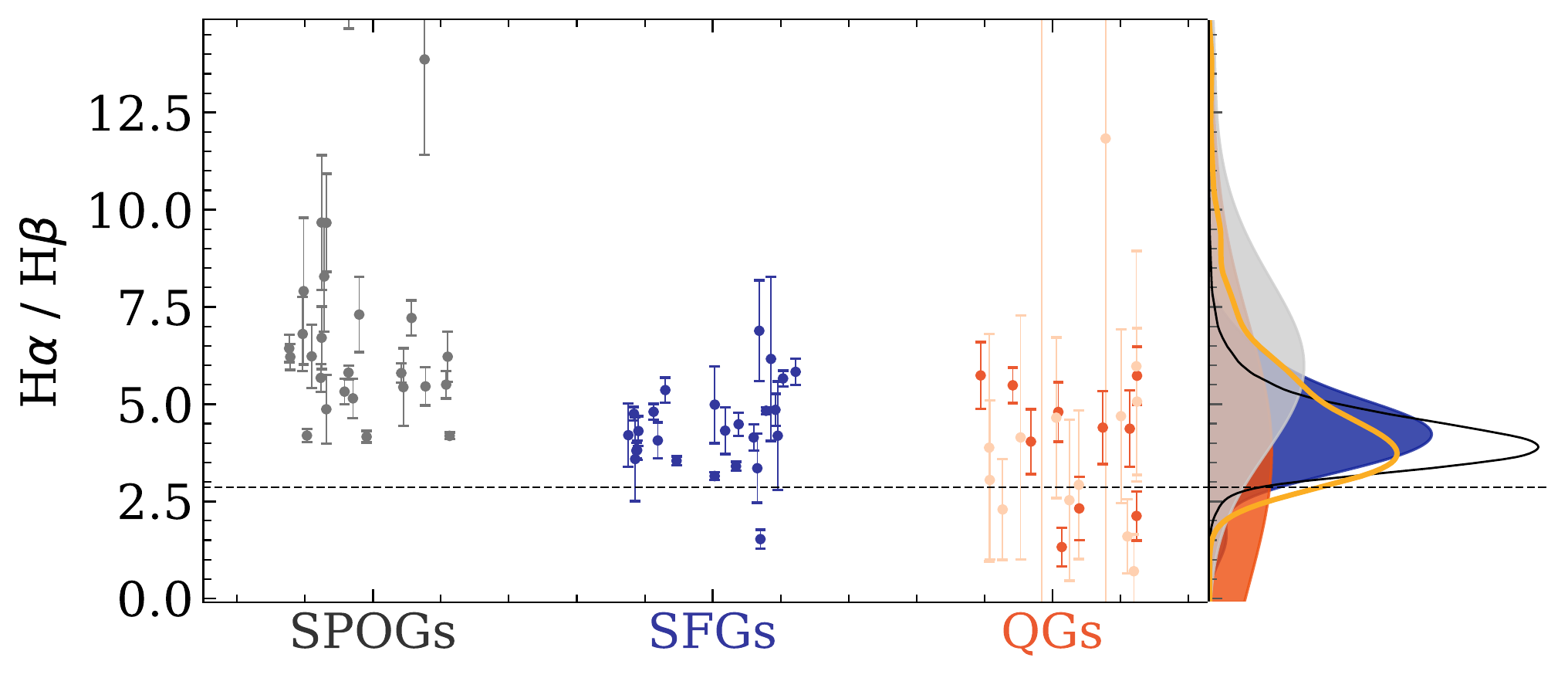}
    \caption{The distributions of the Balmer decrement in HST-SPOGs (grey), star-forming (blue) and quiescent (orange) samples measured with the MPA-JHU emission line catalog. \textbf{Left:} distribution of individual galaxies including H$\alpha$/H$\beta$ uncertainties. Lighter orange points are galaxies with H$\alpha$ or H$\beta$ $S/N < 3$. \textbf{Right:} overall distribution of each sample, smoothed with a Gaussian kernel. Solid black and yellow line shows the distribution for the parent ELG and SPOGs samples respectively. Horizontal dashed line shows H$\alpha$/H$\beta$ = 2.86, typical for star-forming regions \citep{Osterbrock2006}. HST-SPOG sample shows significantly more dust extinction than either calibration and parent samples.}
    \label{fig:dust}
\end{figure}
\begin{figure*}
    \centering
    \includegraphics[width=\linewidth]{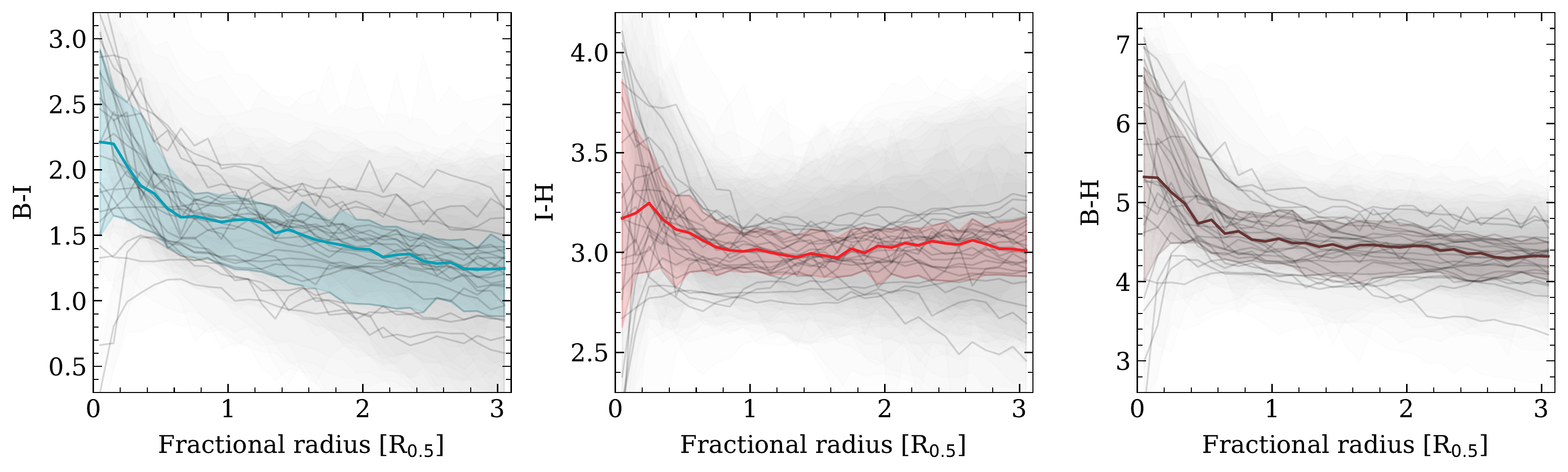}
    \caption{Gradients of \textit{B}-\textit{I} (F438W-F814W, \textbf{left}), \textit{I}-\textit{H} (F814W-F160W, \textbf{center}) and \textit{B}-\textit{H} (F438W-F160W, \textbf{right}) colors of HST-SPOGs computed with \textit{HST} imaging as described in Sec. \ref{sec:color}. Grey solid lines show the color gradient of each HST-SPOG galaxy, normalized by $R_{0.5}$. Grey shaded areas show the 16\ts{th}/84\ts{th} quantiles of each galaxy's color gradient. Colored solid lines and shaded regions show the median and the 16\ts{th}/84\ts{th} gradient quantiles for the entire HST-SPOG sample, respectively. HST-SPOGs generally have negative gradients in the inner $R_{0.5}$, i.e. redder nuclear regions.}
    \label{fig:color}
\end{figure*}
\begin{figure*}
    \centering
    \includegraphics[width=\linewidth]{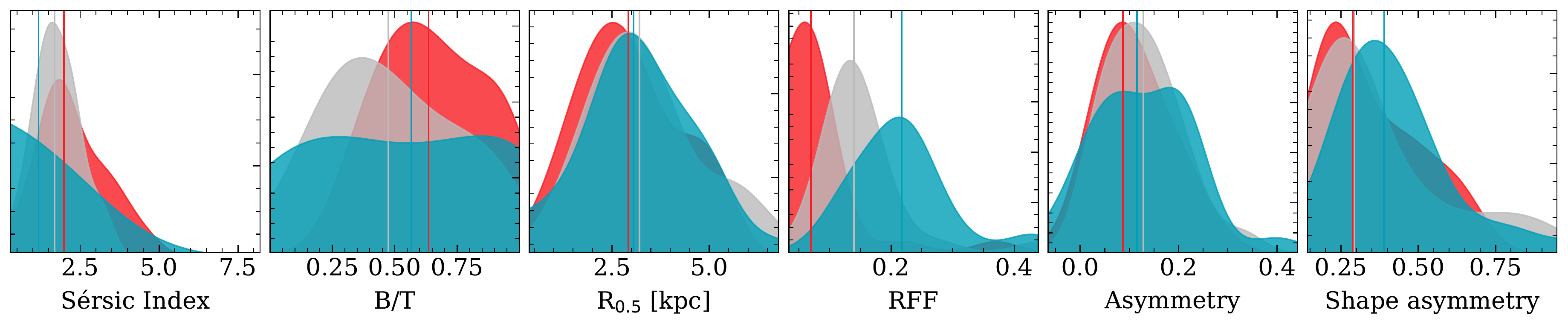}
    \caption{Distributions of morphological parameters for the HST-SPOG sample obtained with \textit{HST} F160W (red), F814W (grey) and F438W (blue) imaging. Left to right, the parameters are: Sérsic index, B/T, Sérsic $R_{0.5}$, RFF, asymmetry and shape asymmetry. HST-SPOGs have slightly more bulge-dominated morphology bluer bands and $>$5$\sigma$ higher RFF in bluer bands.}
    \label{fig:filter_dist}
\end{figure*}
\begin{figure*}
    \centering
    \includegraphics[width=\linewidth]{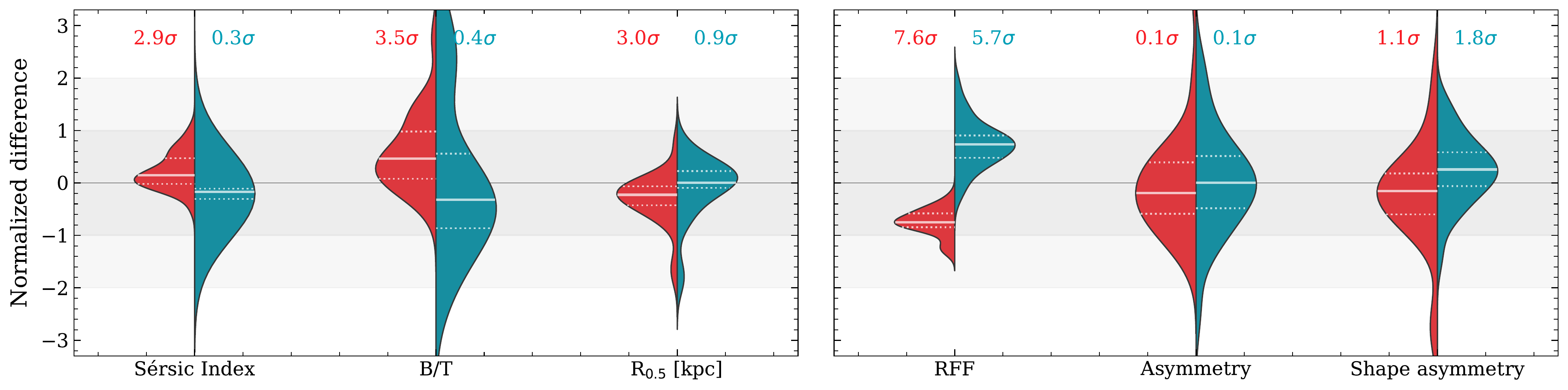}
    \caption{Distribution of \textit{difference} in morphological parameters of HST-SPOGs between \textit{HST} \textit{I} and \textit{H}-band (red) and \textit{I} and \textit{B}-band (blue) imaging. The difference in a parameter $P$ is computed as $P_H - P_I$ and $P_B-P_I$ respectively using Eq. \ref{eq:diff_filt}. White solid and dashed lines show median and 16\ts{th}/84\ts{th} percentiles of each distribution, respectively. Significance levels are shown using a t-test comparison to a 0 mean difference. Higher red (blue) values mean that a parameter is higher in \textit{H} (\textit{B}) imaging than in \textit{I} imaging. \textbf{Left:} difference in structural parameters, concentration, $B/T$ and $R_{0.5}$ showing that HST-SPOGs are more compact and bulge-dominated in redder bands. \textbf{Right:} difference in disturbance parameters RFF, $G-M_{20}$ disturbance and shape asymmetry. HST-SPOGs are $>$5$\sigma$ more disturbed in bluer bands using RFF, but there is no difference in asymmetry or shape asymmetry.}
\label{fig:filter_violins}
\end{figure*}
\pagebreak
\subsection{Dust properties}

To further quantify the dust content of our sample, we looked at the distribution of Balmer decrements in HST-SPOGs compared to the comparison samples. Fig. \ref{fig:dust} shows the comparison of H$\alpha$/H$\beta$ flux measured in the MPA-JHU catalog for HST-SPOGs (grey), star-forming (blue), and quiescent (orange) samples, as well as for all ELG sample galaxies (solid black line). A typical star-forming region has H$\alpha$/H$\beta$ = 2.86, while higher Balmer decrement values indicate dust extinction \citep{Osterbrock2006}. HST-SPOGs indeed show significantly higher Balmer decrements than other galaxies and therefore contain more dust, consistent to our visual inspection of the \textit{HST} snapshots. Since Balmer decrements are obtained with SDSS spectroscopic observations, they are only computed for the inner 3\arcsec of the galaxy.

We then looked at \textit{B}-\textit{I}, \textit{I}-\textit{H} and \textit{B}-\textit{H} color gradients of HST-SPOGs, calculated as described in Sec. \ref{sec:color} and shown in Fig. \ref{fig:color}. Grey solid lines show the average radial color gradient of each HST-SPOG, and light-grey shaded areas show the 18\ts{th}/84\ts{th} quantiles of each gradient. Colored solid lines and shaded areas show the overall average gradient for the HST-SPOG sample and its 16\ts{th}/84\ts{th} quantiles, respectively. 

HST-SPOGs generally have negative color gradients in the inner $R_{0.5}$, indicative of redder nuclei; although some galaxies have positive gradients. There is a much larger variance in \textit{I}-\textit{H} and \textit{B}-\textit{I} gradient in the outer $\sim$2--3$R_{0.5}$: while the average gradient for most galaxies is flat, many galaxies show large scatter in the outer regions. However, this is likely caused by the difference in image depth between the 3 bands.

This result contradicts with previous studies of post-starburst galaxies, which generally show positive color gradients suggestive of outside-in quenching \citep[e.g.,][]{Yang2008, Chen2019}. However, our sample consists of younger post-starbursts which still contain dust. The color gradients may also be caused by a gradient in the dust content rather than in the stellar ages. Although we do not know the dust content in the outer regions of the galaxies, we know that the inner 3\arcsec of HST-SPOGs are more dust-obscured than for average galaxies (Fig. \ref{fig:dust}).

\begin{figure*}
    \centering
    \includegraphics[width=\linewidth]{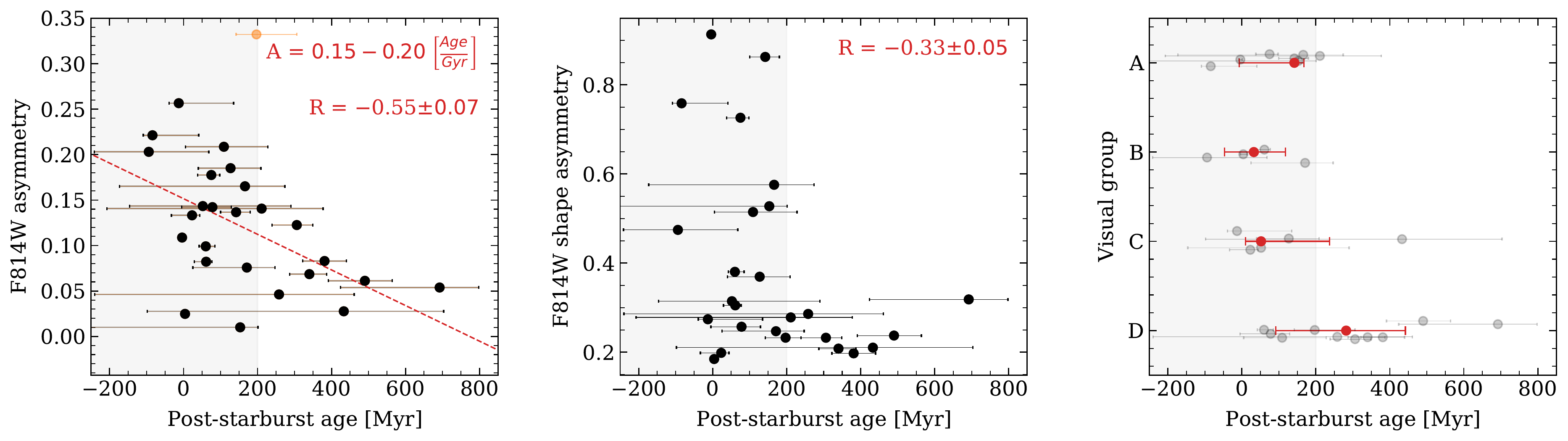}
    \caption{\textbf{Left:} the correlation between \textit{I}-band asymmetry and post-starburst age. Negative ages indicate a burst is ongoing. Red dashed line shows the line of best fit with Pearson $R=-0.55 \pm +0.07$, i.e. linear fit significance of 2.8$\sigma$. The error is given by bootstrapping the age uncertainties. The orange point is the J093820+181953 galaxy with a prominent dust lane, which was excluded as an outlier. Asymmetric features fade over $\sim$750 Myr.  \textbf{Center:} plot of \textit{I}-band shape asymmetry against post-starburst age. There is no clear linear relationship, but only galaxies younger than 200 Myr have high values of $A_S$. \textbf{Right:} distribution of ages in different visual groups from \ref{sec:visual} (grey). The median and 16\ts{th}/84\ts{th} percentiles in each group are shown in red. Galaxies in group D (no clear merger signatures or dust lanes) are, on average, older than any other group. Region with post-burst age $<$200 Myr is shaded in grey in each panel.} 
    \label{fig:age}
\end{figure*}

\subsection{Quantitative Morphology}\label{sec:filt_morph}

As discussed above, HST-SPOGs are significantly dust-obscured, and their visual morphology depends strongly on the imaging wavelength. F160W imaging is infrared and therefore more transparent to dust, leading to fewer dust-absorption features. Moreover, the structure of HST-SPOGs may differ in different bands as they trace different stellar populations. Here, we compare the quantitative morphology in F160W, F438W and F814W bands.

We measured all morphological parameters described in Sec. \ref{sec:morphology} using \textit{HST} imaging. For each parameter in each band, we computed the median of that parameter's distribution in the HST-SPOG sample. These results are shown in Tab. \ref{tab:medians_filters}. We calculated the uncertainty of the median by bootstrapping the sample 1,000 times and finding the 16\ts{th} and the 84\ts{th} percentiles of the bootstrapped medians.

Fig. \ref{fig:filter_dist} shows the distributions of six of these parameters, going left to right: Sérsic index, $B/T$, $R_{0.5}$, RFF, asymmetry, and shape asymmetry. We chose this subset of parameters because it represents well the bulge strength, disturbance and radius of a galaxy and is commonly used in literature. The distributions of the remaining parameters from Sec. \ref{sec:morphology} and the population medians are shown in Appendix \ref{app:more_stats}. The morphology measurements for each individual galaxy in each filter are provided as an online dataset with this paper, and an excerpt of the data is shown in Appendix \ref{app:morphology}.

Each panel in Fig. \ref{fig:filter_dist} shows the distribution of a morphological parameter measured with \textit{H}-band (red), \textit{I}-band (grey) and \textit{B}-band (blue) imaging.  Each distribution is smoothed with a Kernel Density Estimate (KDE) of the underlying data with a bandwidth automatically selected using Scott's rule-of-thumb \citep{Scott1992}. Physical bounds on the KDE were enforced using renormalization (e.g., Sérsic n $>$ 0). Solid red, grey and blue lines show the medians of the corresponding distributions.

HST-SPOGs have intermediate bulge strengths, with average $n = 1.70^{+0.2}_{-0.1}$ and $B/T = 0.47^{+0.03}_{-0.12}$ in \textit{I}-band. They are slightly more bulge-dominated in \textit{H}-band with $n=2.0^{+0.4}_{-0.1}$, and less bulge-dominated in \textit{B}-band with $n=1.2^{+0.4}_{-0.1}$. This indicates that the central bulge is redder, while the disk is bluer, agreeing with the color gradients in Fig. \ref{fig:color}. We note that $B/T$ is actually higher in \textit{B}-band than in \textit{I}-band, contradicting the other results slightly; however, the difference is well within the bootstrapped uncertainty.

HST-SPOGs have relatively high disturbance measurements. Their \textit{I}-band asymmetry, shape asymmetry and RFF are $0.13^{+0.01}_{-0.03}$, $0.29 \pm 0.03$ and $0.14 \pm 0.01$, respectively. While asymmetry and shape asymmetry do not vary much between the bands, RFF depends strongly on wavelength. \textit{H}-band RFF is significantly lower ($0.07^{+0.00}_{-0.01})$ and \textit{B}-band RFF is significantly higher ($0.22^{+0.00}_{-0.01}$). This indicates that disturbances are more apparent in bluer bands.

\renewcommand{\arraystretch}{1.2}
\begin{deluxetable}{r|DDD}
\label{tab:medians_filters}
\tablecaption{Median morphological measurements for HST-SPOGs measured with \textit{HST} \textit{H}-, \textit{I}- and \textit{B}-band imaging.}
\tablehead{
\multicolumn1r{} & \multicolumn2c{F160W} & \multicolumn2c{F814W} & \multicolumn2c{F438W} }
\decimals
\startdata
\multicolumn{7}{l}{Structural} \\
\hline
Sérsic $n$ & 1.98$^{+0.35}_{-0.08}$ & 1.70$^{+0.22}_{-0.12}$ & 1.18$^{+0.41}_{-0.14}$ \\
$B/T$ & 0.64$^{+0.07}_{-0.05}$ & 0.47$^{+0.03}_{-0.12}$ & 0.57$^{+0.09}_{-0.22}$ \\
R$_{0.5}$ [kpc] & 2.92$^{+0.15}_{-0.48}$ & 3.21$^{+0.36}_{-0.33}$ & 3.06$^{+0.46}_{-0.17}$ \\
\hline
\multicolumn{7}{l}{Disturbance} \\
\hline
Asymmetry & 0.09$^{+0.04}_{-0.01}$ & 0.13$^{+0.01}_{-0.03}$ & 0.12$^{+0.05}_{-0.02}$ \\
Shape asymmetry & 0.29$^{+0.05}_{-0.05}$ & 0.29$^{+0.02}_{-0.03}$ & 0.39$^{+0.04}_{-0.04}$ \\
RFF & 0.07$^{+0.00}_{-0.01}$ & 0.14$^{+0.01}_{-0.01}$ & 0.22$^{+0.00}_{-0.01}$ \\
\enddata
\tablecomments{Uncertainties of the median are obtained via bootstrapping each sample 1,000 times.}
\end{deluxetable}


\subsection{Morphology dependence on wavelength}\label{sec:filt_violins}
Population-level analysis is interesting in a statistical sense when characterizing a population; however it is also useful to look at the difference in morphology between filters for each HST-SPOG individually. This can eliminate the scatter due to different morphologies present in our sample and show real wavelength-depending morphology trends, if any.  

We take the \textit{I}-band morphology as the baseline, and look at the difference between the measurements in \textit{I}- and \textit{H}-bands, and \textit{I}- and \textit{B}-bands. We subtract each morphological parameter in \textit{I}-band, $P_I$, from a measurement in a different band, $P_{X}$,  to find the difference $\Delta P_{X-I}$, where $X$ can be either $H$ or $B$. We then look at the distribution of differences in the HST-SPOG sample. We normalize the difference by the overall standard deviation of this parameter in the sample across all wavelengths, $\sigma_{\textrm{all}}$:
\begin{equation}
    \Delta P_{X-I} = \frac{P_X - P_I}{\sigma_{\textrm{all}}}
    \label{eq:diff_filt}
\end{equation}

The violin plots representing this difference are organized in the following way. Figure \ref{fig:filter_violins} shows the distribution of the \textit{difference} in morphology between wavelengths for structural parameters (Sérsic index, $B/T$, $R_{0.5}$; left) and disturbance parameters (RFF, asymmetry, shape asymmetry; right). Each violin consists of two distributions: the difference between $H$- and $I$-band morphology ($\Delta P_{H-I}$, red) and between $B$- and $I$-band morphology  ($\Delta P_{B-I}$, blue). White solid and dashed lines show the median and 16\ts{th}/84\ts{th} quantiles of each distribution, respectively. Values above 0 in red (blue) distributions of a parameter indicate that \textit{H} (\textit{I}) imaging has a higher value of that parameter than \textit{I}-band imaging.

In addition, we calculated the significance of the difference for each set of measurements. We performed a t-test to see if the distribution is consistent with an mean difference of 0.

HST-SPOGs have similar morphology in \textit{B} and \textit{I} bands but are more bulge-dominated in the \textit{H}-band. Moreover, the \textit{H}-band stellar distribution is $3\sigma$ more compact as measured by $R_{0.5}$. As discussed previously, our sample may have a younger stellar disk and an older bulge, resulting in a more disk-like morphology in bluer bands. However, our sample also likely has more central dust, obscuring the young stars in the central region. Even $R_{0.5}$ would not be able to distinguish these two scenarios, as the half-light radius would be lower if the center was less dust-obscured. However, visually HST-SPOGs have a lot of dust substructure in the center (Fig. \ref{fig:thumbnails}), and HST-SPOGs have strong Balmer decrements indicative of dust (Fig. \ref{fig:dust}), supporting the dust origin of the morphological differences.

HST-SPOGs have significantly higher RFF in bluer bands, but their asymmetry is consistent across all bands. Asymmetry is more sensitive to large asymmetric features, such as tidal structures, while RFF measures all deviation of light from a Sérsic distribution. This result implies that bluer imaging contains more internal, symmetric disturbances measured by RFF. On the other hand, global asymmetric features are well-captured by asymmetry in all three bands. High RFF in the blue bands could either be caused by blue knots of star formation or by significant dust obscuration unseen in \textit{H}-band. F160W imaging has a lower spatial resolution, which could lead to a lower RFF. However, F814W and F438W imaging has the same resolution, and RFF is significantly higher in F438W than in F814W, so the wavelength dependence of RFF across all bands is better explained by physical differences in dust or stellar populations than imaging effects.

\subsection{Morphology as a function of Age}

\renewcommand{\arraystretch}{1.2}
\begin{deluxetable*}{r|DDD|DDD}
\label{tab:medians}
\tablecaption{Median morphological measurements for HST-SPOGs and comparison star-forming and quiescent galaxies, calculated with \textit{HST} F814W and SDSS \textit{i}-band.}
\tablehead{
\multicolumn1r{} & \multicolumn6c{\textit{HST} F814W} & \multicolumn6c{SDSS i-band} \\
\multicolumn1r{Parameter} & \multicolumn2c{HST-SPOGs} & \multicolumn2c{SFGs} & \multicolumn2c{QGs} & \multicolumn2c{HST-SPOGs} & \multicolumn2c{SFGs} & \multicolumn2c{QGs}
}
\decimals
\startdata
\multicolumn{13}{l}{Structural parameters} \\
\hline
Sérsic $n$ & 1.70$^{+0.22}_{-0.12}$ & 1.22$^{+0.23}_{-0.11}$ & 2.64$^{+0.34}_{-0.24}$ & 1.43$^{+0.07}_{-0.08}$ & 0.87$^{+0.18}_{-0.09}$ & 2.04$^{+0.03}_{-0.12}$ \\
$B/T$ & 0.47$^{+0.03}_{-0.12}$ & 0.16$^{+0.01}_{-0.02}$ & 0.71$^{+0.02}_{-0.04}$ & 0.17$^{+0.04}_{-0.03}$ & 0.03$^{+0.04}_{-0.01}$ & 0.38$^{+0.01}_{-0.03}$ \\
R$_{0.5}$ [kpc] & 3.21$^{+0.36}_{-0.33}$ & 4.39$^{+0.54}_{-0.35}$ & 2.39$^{+0.22}_{-0.20}$ & 3.92$^{+0.63}_{-0.23}$ & 5.47$^{+0.36}_{-0.34}$ & 3.43$^{+0.26}_{-0.32}$ \\
\hline
\multicolumn{13}{l}{Disturbance parameters} \\
\hline
Asymmetry & 0.13$^{+0.01}_{-0.03}$ & -0.03$^{+0.01}_{-0.01}$ & -0.01$^{+0.01}_{-0.01}$ & 0.03$^{+0.01}_{-0.01}$ & -0.01$^{+0.02}_{-0.03}$ & -0.00$^{+0.01}_{-0.00}$ \\
Shape asymmetry & 0.29$^{+0.02}_{-0.03}$ & 0.27$^{+0.02}_{-0.02}$ & 0.20$^{+0.01}_{-0.01}$ & 0.27$^{+0.02}_{-0.04}$ & 0.30$^{+0.03}_{-0.02}$ & 0.22$^{+0.02}_{-0.00}$ \\
RFF & 0.14$^{+0.01}_{-0.01}$ & 0.07$^{+0.01}_{-0.01}$ & 0.04$^{+0.01}_{-0.01}$ & 0.07$^{+0.01}_{-0.01}$ & 0.07$^{+0.01}_{-0.01}$ & 0.03$^{+0.01}_{-0.00}$ \\
\enddata
\end{deluxetable*}

We investigated the dependence of morphology on the post-starburst age. Fig. \ref{fig:age} shows the post-starburst age (black/grey points) plotted against F814W asymmetry (left), F814W shape asymmetry (center) and visual group (right). The median age within each group is over-plotted in red in the right panel with associated 16\ts{th} and 84\ts{th} percentiles as error bars. Region with post-starburst age $<$200 Myr is shaded in grey.

Although the sample is small, there is a correlation between post-starburst age and asymmetry. For this analysis, we excluded the galaxy J093820+181953, as it is an apparent outlier in Fig. \ref{fig:age} due to its highly obscuring central dust lane. We then fit a straight line to the remaining data and found the correlation to be
\begin{equation}
    A_{\textrm{F814W}} \approx 0.15 - 0.20 \Big[ \frac{\textrm{Age}}{\textrm{Gyr}} \Big],
\end{equation}
indicating that asymmetric features fade on timescales of $\sim$750 Myr, similar to the typical range in post-starburst ages \citep[e.g.,][]{Snyder2011, French2018}. To quantify this correlation, we performed a Pearson's  test by bootstrapping the sample 10,000 times while varying the age measurements within their uncertainties and obtained $R = -0.55 \pm 0.007$. This corresponds to $p\approx0.004$ or a 2.8$\sigma$ significance. This correlation is marginally significant, as we are limited by the sample size and large age uncertainties, but much more significant than that found in \cite{Pawlik2016} using SDSS imaging. 

Similarly, we looked at the relationship between shape asymmetry and post-starburst age. There is no clear linear trend, and Pearson correlation is $R = -0.32 \pm 0.05$, only 1.6$\sigma$ significant. However, it is apparent that only galaxies with $<$200 Myr post-burst age show high shape asymmetry and hence likely display tidal features. Previous simulations by \cite{Pawlik2018} and \cite{Snyder2015a} similarly show tidal merger signatures fading on the timescales of $\sim$200--500 Myr, agreeing with our results. 


In terms of our visual classification, there is no age difference between groups A, B and C; group D galaxies are significantly older. Galaxies in groups A, B and C are almost all younger than 200 Myr; group D galaxies are generally older than 200 Myr. This distinction implies that, although groups A, B and C appear visually different, they trace similar evolutionary stages of a post-starburst galaxy. On the other hand, group D contains only older post-starbursts. Group D galaxies were visually selected as those without clear disk morphology or disturbances, agreeing with the paradigm that low-redshift PSBs transition towards early-type morphologies.


\section{Results: HST-SPOGs vs comparison galaxies}

We compared the morphology of HST-SPOGs to that of the comparison SFG and QG samples. As stated in Section \ref{sec:data}, we looked at the morphology of all galaxies in \textit{HST} F814W imaging and corresponding SDSS \textit{i}-band imaging. This section lists the numerical results of our analysis, while Sec. \ref{sec:discussion} discusses our important findings and their scientific implications.

\begin{figure*}
    \centering
     \includegraphics[width=\linewidth]{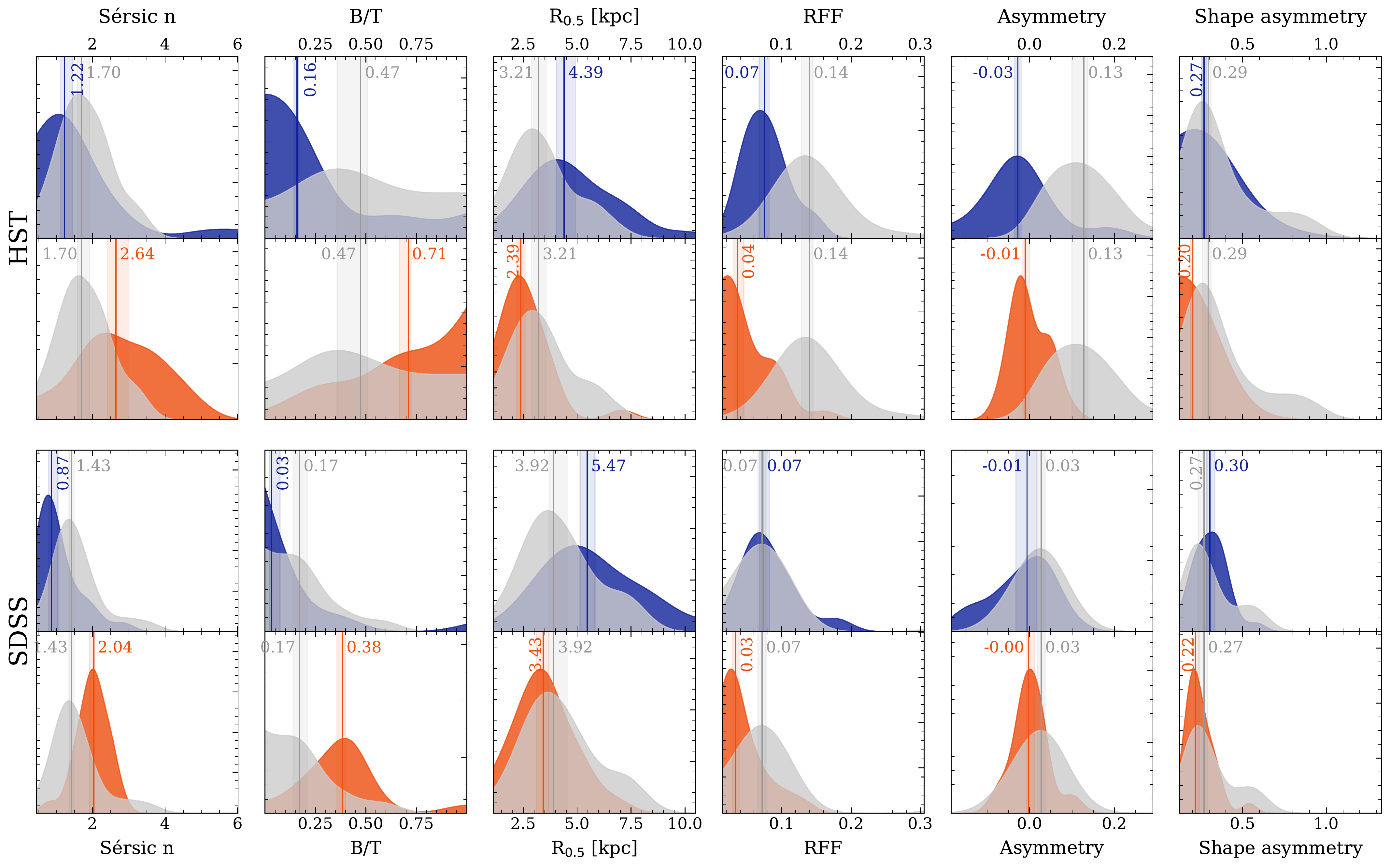}
    \caption{Distributions of morphological parameters for 3 samples: HST-SPOGs (grey),  star-forming galaxies (blue), and quiescent galaxies (orange). Left to right, the parameters are: Sérsic index, $B/T$, asymmetry, RFF, and Sérsic $R_{0.5}$.  Parameters were computed using \textit{HST} F814W imaging (top) and SDSS \textit{i}-band imaging (bottom). HST-SPOGs are morphologically similar to star-forming galaxies when using SDSS imaging but show significantly more asymmetry and RFF in \textit{HST}.} 
    \label{fig:morph_distributions}
\end{figure*}
\begin{figure*}
    \centering
    \includegraphics[width=\linewidth]{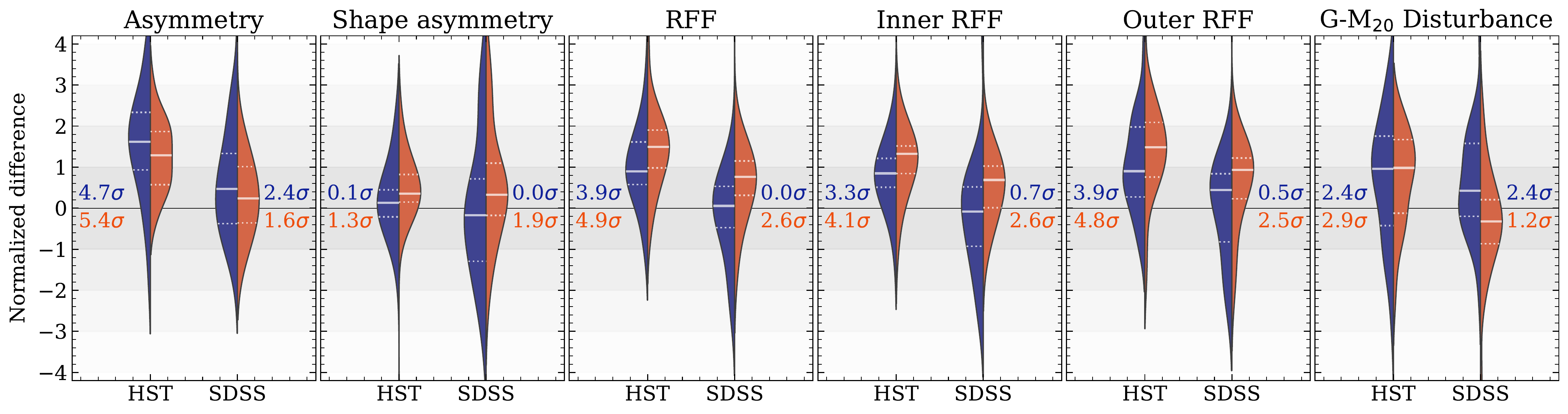}
    \caption{Distributions of \textit{difference} in disturbance metrics of HST-SPOGs and comparison star-forming (blue) and quiescent (red) galaxies, as defined in Eq. \ref{eq:diff}. Each panel shows the different distribution for a metric obtained with \textit{HST} F814W imaging (left) and SDSS \textit{i}-band imaging (right). Higher values in the blue (red) distribution mean that the HST-SPOG has higher disturbance values than the matched comparison SFG (QG). The white ticks in each violin show the median and $\pm 1 \sigma$ of each distribution. Significance of the difference is calculated using a t-test comparison to a 0 mean difference. Using \textit{HST} imaging, HST-SPOGs are more than 3$\sigma$ more disturbed using most metrics. The difference in $G-M_{20}$ is marginally significant, and there is no difference in shape asymmetry. In SDSS, no parameter shows a significant difference.} 
    \label{fig:a_violins}
\end{figure*}
\subsection{Population-level morphology}\label{sec:populations}

First, we looked at the morphological differences between HST-SPOGs, star-forming, and quiescent galaxies on the population-level, similar to Sec. \ref{sec:filt_morph}. For this analysis, we had 6 data samples: \textit{HST} and SDSS imaging of HST-SPOGs, SFGs, and QGs. For each morphological parameter from Sec. \ref{sec:morphology}, we computed the median and the 16\ts{th}/84\ts{th} percentiles of that parameter's distribution in each sample. 

Fig. \ref{fig:morph_distributions} shows the distributions of these parameters, going left to right: Sérsic index, $B/T$, $R_{0.5}$, RFF, asymmetry, and shape asymmetry. The medians of these parameters in each sample are given in Tab. \ref{tab:medians}. The results for the remaining parameters described in Sec. \ref{sec:morphology} are shown in Appendix \ref{app:more_stats}. The morphology measurements for each galaxy from each sample are provided as an online dataset with this paper.

The top half of Fig. \ref{fig:morph_distributions} shows the results using the \textit{HST} F814W imaging, and the bottom half -- SDSS \textit{i}-band. In each half, the top panel shows the distribution of HST-SPOGs (grey) and SFGs (blue); the bottom panel shows HST-SPOGs (grey) and QGs (orange). As in Sec. \ref{sec:filt_morph}, each distribution is smoothed with a Gaussian KDE. The median of the distribution is shown as a labelled solid line, and the uncertainty of the median is shown as a shaded region around the median.

HST-SPOGs have intermediate bulge strengths, with average $n = 1.70^{+0.2}_{-0.1}$ and $B/T = 0.47^{+0.03}_{-0.12}$ in \textit{HST} F814W imaging. This is between our star-forming and quiescent samples and consistent with other similar results by \cite{Blake2004, Pawlik2016, Chen2019}. 


Both in SDSS and \textit{HST} imaging, the Sérsic index of post-starbursts is more similar to star-forming galaxies, indicating a stronger disk component. Interestingly, in \textit{HST} imaging, the average $B/T$ of our sample is closer to that of quiescent galaxies, and the effective radius in both datasets is also close to that of QGs. This indicates that although the overall morphology of HST-SPOGs may be disk-like, they host compact central bulges.


Finally, HST-SPOGs have significantly higher asymmetry ($0.13^{+0.01}_{-0.03})$ and RFF ($0.14\pm0.01$) than either comparison sample when observed with \textit{HST}. This indicates that HST-SPOGs are more disturbed than \textit{regular} galaxies, although by construction of our comparison sample, we cannot explicitly state that HST-SPOGs are more disturbed than \textit{field} galaxies since we excluded mergers. When observed with SDSS, HST-SPOGs are more disturbed than QGs but not enough to distinguish them from regular SFGs. However, the distribution of shape asymmetry in HST-SPOGs is consistent with that of SFGs in both \textit{HST} and SDSS imaging. The implications of this discrepancy are discussed in Sec. \ref{sec:mergers}.

\subsection{Morphology of matched galaxies}\label{sec:violins}


Similarly to Sec. \ref{sec:filt_violins}, we then looked at the difference in morphology between each HST-SPOG and a matched comparison galaxy. Morphological measurements depend on spatial resolution \citep[e.g.,][]{Lotz2004}. Lower-mass galaxies are smaller, and therefore identical imaging resolves larger physical scales in these galaxies. We eliminate this bias by comparing each HST-SPOG to a mass- and redshift-matched comparison galaxy.

For each morphological parameter $P$, we subtract the comparison measurement, $P_{\textrm{comp.}}$, from the HST-SPOG measurement, $P_{\textrm{HST-SPOG}}$,  to find the difference $\Delta P$. We then look at the distribution of differences in all samples. For each parameter and instrument sample, we normalize the difference by the overall standard deviation of this parameter in the entire dataset, $\sigma_{\textrm{all}}$:

\begin{equation}
    \Delta P = \frac{P_{\textrm{HST-SPOG}} - P_{\textrm{comp.}}}{\sigma_{\textrm{all}}}
    \label{eq:diff}
\end{equation}

The violin plots in the following subsections are organized in the following way. Figures \ref{fig:a_violins} and \ref{fig:bulge_violins} show the distribution of \textit{difference} $\Delta P$ between HST-SPOGs and matched comparison galaxies for disturbance and structural parameters respectively. 

Each panel contains two violins, showing the distribution of differences between HST-SPOGs and comparison galaxies with \textit{HST} data (left) and SDSS data (right). Each violin consists of two halves: the difference between HST-SPOGs and SFGs (blue, left half) and HST-SPOGs and QGs (orange, right half). We also calculated the significance of the difference for each set of measurements by performing a t-test, same as in Sec. \ref{sec:filt_morph}.

\subsubsection{Disturbance}\label{sec:a_violins}

\begin{table}
    \caption{Fraction of HST-SPOGs that are more disturbed than their matched comparison galaxies}
    \centering
    \begin{tabular}{rcc|cc|cc}
    \hline \hline
         & \multicolumn{2}{c}{SFGs} & \multicolumn{2}{c}{QGs} &
         \multicolumn{2}{c}{Both} \\
        Parameter & \textit{HST} & SDSS & \textit{HST} & SDSS & \textit{HST} & SDSS \\
        \hline
        Asymmetry & 85\% & 65\% & 92\% & 62\% & 85\% & 50\% \\
        Shape asymmetry & 62\% & 42\% & 88\% & 73\% & 58\% & 42\% \\
        RFF & 88\% & 54\% & 92\% & 81\% & 88\% & 50\% \\
        RFF\sub{in} & 92\% & 46\% & 92\% & 73\% & 88\% & 42\% \\
        RFF\sub{out} & 85\% & 65\% & 92\% & 81\% & 81\% & 58\% \\
        $G/M_{20}$ disturbance & 73\% & 65\% & 69\% & 35\% & 58\% & 23\% \\
        \hline
    \end{tabular}
    \label{tab:disturbance}
\end{table}

Fig. \ref{fig:a_violins} shows the distributions of the difference in disturbance between HST-SPOGs and comparison galaxies. Left to right, the panels show 6 measurements of disturbance: asymmetry, shape asymmetry, RFF, RFF\sub{in}, RFF\sub{out} and G-M$_{20}$ disturbance.

In \textit{HST} data, HST-SPOGs are more disturbed than SFGs and QGs at a higher than $3\sigma$ level using most parameters except G-M$_{20}$, disturbance and shape asymmetry. On the other hand, no parameters show a significant difference in SDSS imaging. We computed the fraction of galaxies that are more disturbed than their comparison counterpart, using \textit{HST} and SDSS imaging, shown in Tab. \ref{tab:disturbance}. We find that 85\%--88\% HST-SPOGs are significantly more disturbed than comparison galaxies; however, this disturbance is only seen with high-resolution \textit{HST} imaging and not with SDSS.

On the other hand, only 65\% of HST-SPOGs have a higher shape asymmetry than both comparison samples with \textit{HST}. This is a relatively small fraction, considering 50\% would mean that two compared samples are consistent with each other. Although the fraction of HST-SPOGs with high $A_S$ in \textit{HST} imaging increased compared to 42\% with SDSS imaging, this increase is smaller than for other disturbance parameters. This suggests that the increased resolution highlights \textit{internal} disruptions, rather than obvious disturbances to a galaxy's shape such as tidal tails. The implications of this to the origin of disturbances in HST-SPOGs are discussed in Sec. \ref{sec:mergers}.




Finally, HST-SPOGs have only a marginally higher G-M$_{20}$ disturbance than comparison star-forming galaxies. G-M$_{20}$ is particularly sensitive to late-stage merger features such as double nuclei and enveloped companions, leading to high Gini and low M$_{20}$ values. The efficacy of the G-M$_{20}$ metric on other disturbances, such as dust structures, was never tested, but the metric was not designed to detect them. Therefore, relatively normal G-$M_{20}$ values could be another indicator that the disturbance features are unlikely directly merger-induced.


\pagebreak
\subsubsection{Bulge strength \& Radius}

\begin{figure}
    \centering
    \includegraphics[width=\linewidth]{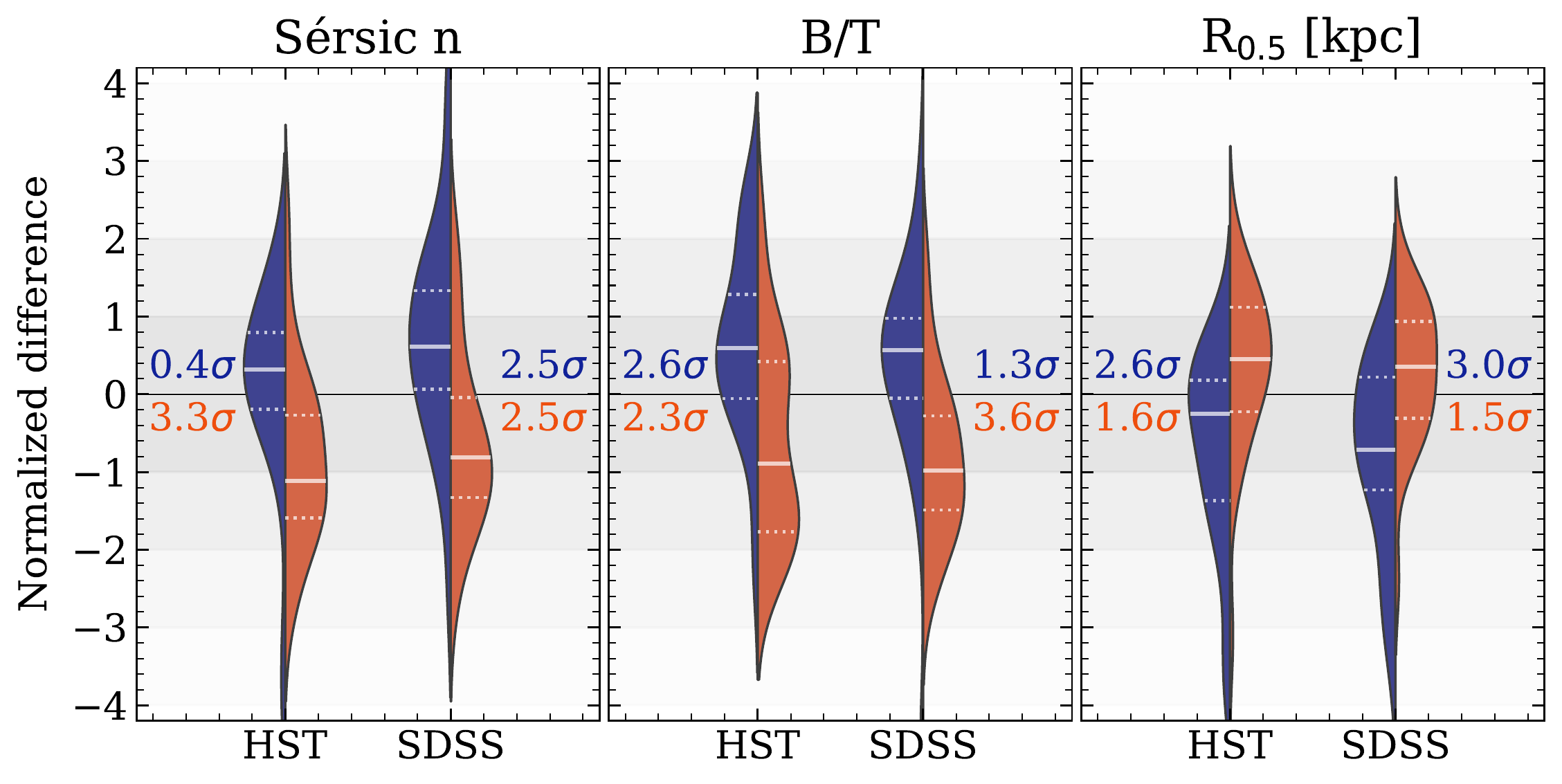}
    \caption{Same as Fig. \ref{fig:a_violins}, showing the distribution of difference in $n$, $B/T$, and $R_{0.5}$ between HST-SPOGs, QGs and SFGs. The comparison samples show an offset from each other, with quiescent galaxies having higher bulge strength in all parameters. HST-SPOGs lie in the intermediate morphology range and therefore show little offset from either sample, with slight preference towards disk-like morphologies seen in \textit{HST} imaging.
} 
    \label{fig:bulge_violins}
\end{figure}

Fig. \ref{fig:bulge_violins} shows the difference in overall morphology between matched HST-SPOGs and comparison galaxies, similar to Sec. \ref{sec:a_violins}. We computed the difference for (left to right) Sérsic index $n$, bulge-to-total light ratio $B/T$, and half-light radius $R_{50}$. The other bulge strength parameters show the same trends and are omitted. As expected, there is a clear offset between star-forming galaxies (blue) and quiescent galaxies (orange) in terms of bulge strength and radius. Quiescent galaxies have stronger bulges and are more compact than their star-forming counterparts. HST-SPOGs, as before, have intermediate morphology between the two comparison samples. As we discussed in Sec. \ref{sec:populations}, HST-SPOGs have disk-like Sérsic indices (with a $3.3\sigma$ deviation from QGs in \textit{HST}) and disk-like $B/T$ in SDSS imaging but not in \textit{HST} data. However, HST-SPOGs are more compact than SFGs at $2.6\sigma$ in \textit{HST} imaging and $3.0\sigma$ in SDSS imaging. The implications of their disk-like morphology with compact cores are discussed in Sec. \ref{sec:trans}.

\section{Discussion}\label{sec:discussion}

\subsection{Transitioning from late- to early-type}\label{sec:trans}

In both Fig. \ref{fig:morph_distributions} and Fig. \ref{fig:bulge_violins}, we see that HST-SPOGs have an intermediate bulge strength between SFGs and QGs but are not significantly different from either sample, agreeing with previous studies \citep{Blake2004, Pawlik2016, Chen2019}. This suggests that HST-SPOGs are in the middle of their morphological transformation towards early-type galaxies, and therefore that this transformation occurs on the same timescales as the post-starburst phase. Moreover, in Fig. \ref{fig:age}, we found that galaxies visually identified as mergers or disks have post-burst ages $<$200 Myr (Fig. \ref{fig:age}, while galaxies with spheroidal morphologies are older than 200 Myr, providing another evidence for a morphological transition over the post-starburst lifetime. 

Subtle differences in bulge strength metrics provide an interesting insight into the evolution of HST-SPOGs. Their Sérsic indices are more consistent with that of star-forming galaxies, but their compactness and $B/T$ and $R_{0.5}$ are slightly more similar to that of quiescent galaxies. This difference indicates that HST-SPOGs may have a generally disky morphology with a bright, compact core, perhaps indicative of an ongoing starburst or an AGN.

In contrast, some other studies founnd that post-starbursts are more compact than QGs. \cite{Almaini2017} recently showed that high-redshift PSBs are extremely compact, indicative of a quenching mechanism that drives the gas inwards and compactifies the galaxy. However, this discrepancy is unsurprising: high-redshift quenching events are expected to be more violent: galaxies are more gas-rich at $z > 1$ \citep[e.g.,][]{Daddi2010}, leading to highly dissipative wet mergers and disk instabilities that are extremely rare in the local Universe \citep{Hopkins2009, Dekel2013}. \cite{Yang2008} found that HST-observed E+A galaxies have extremely high Sérsic indices, disagreeing with our results. However, they do not include a PSF component in their \textsc{GALFIT} model. \textit{HST} imaging provides a sufficiently high resolution that the galaxy nucleus acts as a point source in most galaxies and needs to be fit with a separate component to obtain an accurate model. Without a PSF component, we obtain similarly high Sérsic indices for HST-SPOGs, as well as for comparison star-forming and quiescent galaxies, showing that these high Sérsic indices are unrealistic. 

\subsection{88\% of HST-SPOGs have fading disturbances}\label{sec:disturbances}

In this work, we found that 88\% of our HST-observed post-starburst galaxies are more disturbed than their comparison \textit{regular} star-forming or quiescent galaxies. It is difficult to make a direct comparison of our sample to the general field galaxies, since the techniques of identifying disturbed or merging galaxies differ across studies. Searches of close companions find much lower merger fractions in the field \citep[2--5\%; e.g.,][]{Ellison2008,Patton2011,Man2016}. Major merger fractions identified visually or using asymmetry in SDSS imaging are similar \citep[1--5\%;][]{Darg2010,Casteels2014}. Galaxy Zoo 2 classifications, not aimed specifically to detect mergers, show that $\sim$20\% of local (z$<$0.03) galaxies have ``odd features" \citep[rings, lenses, dust lanes, tidal features, and other irregularities;][]{Willett2013}, similar to those we detected in our work with asymmetry and RFF. These local objects are better resolved with SDSS, so they offer a reasonably good comparison to our analysis using higher-resolution \textit{HST} imaging of more distant galaxies. The fraction of HST-SPOGs with detectable disturbances is still higher than the field average found by \cite{Willett2013}. Overall, although a direct comparison to other studies is challenging, HST-SPOGs are clearly more disturbed than the general field population.


The disturbed fraction of HST-SPOGs is also significantly higher than the 30\% - 50\% reported by previous studies of post-starburst galaxies \citep{Blake2004, Goto2005, Pawlik2016, Yang2008}. However, all of these studies except \cite{Yang2008} were based on ground-based imaging from SDSS or older surveys. When using only SDSS \textit{i}-band data, we similarly found that approximately 50\% of HST-SPOGs are more disturbed than their comparison matches, while the other 50\% are not. Therefore, it is very likely that the majority of post-starburst galaxies have disturbed morphologies, but many of these disturbances remain undetected by low-resolution ground-based surveys like SDSS. The disturbance features have smaller scales that are washed away by the atmospheric seeing and require higher-resolution imaging to detect.

The only study that used space-based imaging, \cite{Yang2008}, analyzed 21 E+A galaxies and found that 11 have dramatic tidal features, leading to a 50\% merger fraction, lower than the disturbance fraction we found. However, other galaxies in this sample exhibited disturbances such as bars and dust lanes and showed significant Sérsic residuals. Although these galaxies were not classified as mergers in \cite{Yang2008}, they would likely be classified as disturbed in our analysis as long as they are more disturbed than similar comparison galaxies. 

We also found a correlation between asymmetry and a post-starburst age: only post-starbursts with post-burst age $\lessapprox 200$ Myr have a high shape asymmetry ($A_S > 0.3$), although most (85\%) have a higher asymmetry than their matched comparison galaxies. An approximate linear regression showed that asymmetry will fade completely on timescales of $\sim$750 Myr, comparable to the post-starburst phase lifetime \citep[e.g.,][]{Snyder2011, French2018}. Although the significance for this correlation is marginal (2.8$\sigma$), likely due to large uncertainties in post-starburst age estimates and a small sample size, it is much higher than that found in \cite{Pawlik2016} using SDSS imaging. 

The correlation between asymmetry and the post-burst age can also explain the discrepancy between the disturbed fraction between our sample and that of other studies. Our sample is biased towards young post-starbursts: the SPOG sample is already younger on average than other post-starburst selections, such as E+As \citep{spogs, French2018}. HST-SPOGs are expected to be even younger as they are more gas- and dust-rich than the parent SPOG sample. Therefore, since we see that asymmetry fades over time, we expect other selections of older post-starbursts to have lower disturbed fractions than HST-SPOGs. This is especially important if dust is indeed the cause of the high RFF we observed, since dust content decreases with post-starburst age \citep[e.g.,][]{Smercina2018}.



\subsection{Are post-starbursts also post-mergers?}\label{sec:mergers}

The possible merger origin of post-starburst galaxies has been long debated in the field. Gas-rich mergers are an effective way to cause a starburst and subsequently quench star formation \citep[e.g.,][]{Bekki2005} and are an important formation mechanism for post-starburst galaxies in simulations \citep{Zheng2020, Lotz2020}. Tidal features specifically are extremely common amongst post-starburst galaxies \citep[e.g., 52\% in][]{Yang2008}. Outside of clusters, tidal features can only be induced by galaxy interactions, so therefore mergers are a likely formation pathway in at least half of post-starbursts. However, the origin of the remaining half has historically been an unanswered question.

In our study, we evaluate the \textit{disturbance} of HST-SPOGs, which is not equivalent to identifying merger signatures. Measures such as asymmetry and RFF simply detect deviations from a smooth stellar distribution. Such deviations can be caused by merger-induced tidal features but could also be due to spiral features, dust lanes, bars, etc. Recent works show that machine learning-based approaches are better at specifically identifying galaxy mergers \citep[e.g.,][]{Snyder2019, Ferreira2020}. One morphology metric that less ambiguously detects post-mergers is shape asymmetry \citep{Pawlik2016}, which looks for deviations in a galaxy shape rather than light distribution and detects asymmetric tidal features, but detecting symmetric merger signatures without machine learning methods remains a challenge.

While 88\% of HST-SPOGs are more disturbed than comparison galaxies, only 65\% have a higher shape asymmetry. Therefore, our sample is clearly highly disturbed, but the disturbances are not necessarily caused by tidal perturbations. Instead, we see that the disturbances are best captured by metrics of internal structure, such as RFF or asymmetry. Our results indicate that about 30\% of HST-SPOGs are internally disturbed without significant tidal features.  Similar results have been found in morphological studies of low-redshift AGN-hosts. The majority of AGN-hosts do not appear to have a merger origin \citep[e.g.,][]{Grogin2005,Cisternas2011, Kocevski2012} but tend to have chaotic or spiral nuclear dust structures \citep{Malkan1998,Deo2006,SimoesLopes2007} and a higher prevalence of inner rings \citep{Hunt1999}.

We also see that the RFF disturbance is significantly higher in bluer imaging at a $> 5\sigma$ level, while shape asymmetry is similar in all bands. To the first order, tidal perturbations are gravitational and affect all stellar populations equally; so consistent shape asymmetry across bands is expected. A higher RFF in blue imaging can be explained by either a strongly perturbed young stellar population or by a high degree of dust obscuration. However, an extreme difference in stellar populations is required to explain the $5\sigma$ RFF difference between \textit{B} and \textit{I} bands. On the other hand, HST-SPOGs contain significant amount of dust (Fig. \ref{fig:dust}), which appears disturbed upon visual inspection of galaxy snapshots. Therefore, it is likely that the high internal disturbance is caused by dust structures, similar to AGN hosts. This indicates that the disturbances we see are caused primarily by dust substructure, rather than merger-induced tidal features.



There are many processes that could lead to a disturbed dust morphology and quenching of star formation that are not necessarily associated with mergers. For example, outflows from a starburst- or AGN-driven feedback could lead to the disruption of dust, dusty outflows and dust cones \citep[e.g.,][]{Obied2016, Kaufman2020}, while accreting AGNs are linked to nuclear dust spirals \citep[e.g.,][]{Martini2003}. HST-SPOGs still contain large reservoirs of molecular gas and dust. \cite{spogs2} showed that the parent sample of HST-SPOGs has a significant NaD excess, indicating possible ongoing outflows. As SPOGs transition into early-type, they have to lose their molecular gas, so outflows are necessary at some point in the post-starburst evolution. Outflows can also inject turbulence into the interstellar medium, leading to the suppression of star formation before the gas is expelled \citep{Nesvadba2010, Alatalo2015,Lanz2016}. It is therefore plausible that we are seeing structural disturbances in a galaxy's dust content caused by these outflows.

However, these disturbances could also be of a merger origin, even though there are no strong tidal features in $\sim$30\% of our sample. HST-SPOGs are \textit{post}-starbursts, so if they are quenching after a merger, then it must have already happened, triggered a starburst, and started fading. Simulations find that asymmetric features fade on timescales between $\sim$200-500 Myr \citep[e.g.,][]{Pawlik2018,Snyder2015a}, although the lifetime of \textit{detectable} disturbances depends strongly on merger mass, gas and dust fraction, orbital arrangement of merging galaxies, and even the way mock observations were created \citep[e.g.,][]{Lotz2008, Pawlik2018, Snyder2015a}, making the comparison to simulations challenging.

We found that asymmetry and shape asymmetry decrease with post-starburst age on similar timescales to simulations. Only HST-SPOGs younger than 200 Myr have $A_S > 0.3$. The decline in asymmetry is smoother, and it takes $\sim$750 Myr for asymmetry to fade entirely. The similarity in asymmetry fading timescales between our sample and simulations suggests that a prior merger is a plausible origin of the disturbances we observed. HST-SPOGs that have higher RFF but lower shape asymmetry than comparison galaxies are circled in Fig. \ref{fig:age}. They appear to preferentially be slightly older, supporting the idea that perhaps obvious post-merger tidal features have faded, while internal disturbances have not. However, we cannot conclude this with high significance due to our small sample size. We plan to follow up this study with a new \textit{HST} program, obtaining imaging of a large subset of the SPOGs sample, spanning a range of stellar masses, redshifts, post-burst ages, and environments. Deeper observations with \textit{HST}, and eventually the Vera Rubin Observatory, will enable detecting fainter tidal features and thereby constraining the merger origin of post-starbursts even further.

If the majority of HST-SPOGs are indeed post-mergers, then our results suggest that disturbances in dust and molecular gas structure fade slower than disturbances in stellar structure. It is possible that a merger simply triggers a different instability leading to a disturbed internal structure \citep[e.g., turbulence, bars, AGN, or starburst,][]{Bekki2005,Wild2009,Snyder2011}. 

Disturbance of the dust substructure is particularly challenging to compare to cosmological simulations, as many simulations lack the necessary spatial resolution to resolve cold gas that is correlated with dust \citep[e.g., $\sim$1 kpc per resolution element in Illustris,][]{Nelson2015}. Finally, while galaxy mergers themselves are easy to simulate, the effect of the merger on the galaxy post-coalescence and the evolution of the post-starburst phase is notoriously difficult to reproduce due to the complex interplay of starburst- and AGN-feedback and shocks, which all require poorly-constrained sub-grid routines. In particular, \cite{Zanisi2021} show that Illustris simulations fail to re-create small-scale morphological features, important in our sample, in compact and quenched galaxies. Isolated merger simulations, tailored to provide high enough resolution to reproduce post-starburst properties, can provide a better morphological comparison \citep[e.g.,][]{Zheng2020, Lotz2020}. However, new-generation cosmological simulations may also reach the necessary spatial resolution alongside a realistic set of sub-grid routines to simulate the post-starburst phase well enough to enable studying this population \citep[e.g.,][]{firebox_prop,Simons2019,Nelson2019}.

\subsubsection{Inside-out quenching or dust?}

We found that HST-SPOGs have more bulge-dominated structure in \textit{H}-band and more disk-like structure in \textit{B}-band at a $3\sigma$ level, corresponding to a red central bulge and a blue disk. Our sample is also, on average, more compact in \textit{H}-band than in \textit{B}-band. Finally, the radial color gradients seen in Fig. \ref{fig:color} indicate that HST-SPOGs have red nuclei, but flat color gradients outside of the inner $R_{0.5}$.

Red bulges in HST-SPOGs could indicate inside-out quenching, where the central bulge is quenched first, followed by a slower quenching of the disk. However, multiple other studies find that post-starburst galaxies generally have an opposite color gradient: bluer cores and redder disks; although some scatter exists across different post-starburst samples \citep{Yang2008, Chen2019}. Moreover, \cite{FraserMcKelvie2018} find that intermediate-mass early-type galaxies, likely descendants of post-starbursts, have younger central bulges suggestive of outside-in quenching, and \cite{Zheng2020} found evidence for outside-in quenching in simulated post-starbursts.  Finally, \cite{Vika2015} found that regular star-forming galaxies also generally have higher bulge strengths in redder bands, so our findings may be indicative of a typical trend for late-type galaxies rather than a hint of inside-out quenching.


On the other hand, the observed red bulges can be caused by dust reddening in the center. HST-SPOGs have large amounts of dust in their central 1.5\arcsec regions (Fig. \ref{fig:dust}, i.e. in their inner $\sim$0.75R$_{0.5}$. Dust preferentially obscures stellar light in \textit{B} and \textit{I} bands, so the central regions of HST-SPOGs are definitely reddened by dust. With this data alone, we cannot conclude that the observed color and morphology gradients are definitely caused by dust, since we do not know whether the dust content decreases at larger radii. However, \cite{Smercina2018} show that molecular gas and dust of post-starburst galaxies is generally constrained to their nuclear regions. Considering this and the fact that 1) many other studies suggest that post-starbursts have bluer cores and 2) our sample contains younger, more gas- and dust-rich post-starbursts, then it is very likely that the observed color gradient is caused by a gradient in the galaxies' dust content. This study paves the way for the future observations with the \textit{James Webb Space Telescope}, which will enable robustly breaking the age-dust degeneracy with high-resolution imaging of the dusty regions in the MIR.


\section{Conclusion}

Previous ground-based studies of post-starburst galaxies found that about 50\% of PSBs are morphologically disturbed, and likely formed as a result of a merger \citep[e.g.,][]{Blake2004, Yang2008, Pawlik2016}. So how did the remaining 50\% quench?

We have performed a detailed morphological study of 26 shocked, CO-detected post-starburst galaxies (HST-SPOGs) with new \textit{HST} imaging in F438W, F814W and F160W bands. We found the following.

\begin{enumerate}[itemsep=0mm]
    \item HST-SPOGs have an intermediate morphology between late- and early-type galaxies.
    \item HST-SPOGs have redder bulges and bluer disks. 
    \item 88\% of our sample has a higher RFF than a mass- and redshift-matched comparison sample of star-forming and quiescent galaxies, and hence is more structurally disturbed.
    \item Only $\sim$50\% of our sample appears more disturbed than the comparison samples in lower-resolution SDSS imaging.
    \item The disturbances are of the \textit{internal structure}, as traced by asymmetry and RFF, rather than shape of the galaxy, as traces by shape asymmetry, so high-resolution imaging is essential to identify them.
    \item These disturbances are most prominent in \textit{B} and \textit{I} bands and not in \textit{H}-band, and hence likely caused by a dust-obscured central structures.
    \item These disturbances fade over time: they become less prominent after $\sim$200 Myr and fade entirely after $\sim$750 Myr. Our sample contained preferentially younger post-starbursts, so older post-starburst samples (such as E+As) are likely to show fewer disturbances.
\end{enumerate}

The intermediate morphology of HST-SPOGs suggests that they are currently transitioning from late- to early-type galaxies alongside their transition into quiescence. The color gradient we found disagrees with color gradients found in similar studies \citep{Yang2008, Chen2019}. This could indicate an inside-out quenching mechanism; but since our sample is relatively young, gas- and dust-rich, it is likely that the apparent gradient is caused by central dust obscuration in bluer bands. 

We conclude that the majority of post-starbursts are disturbed. These disturbances are small-scale and are unresolved by SDSS imaging and are likely caused by a highly disrupted dust structure in our galaxies.  

These small-scale disturbances do not necessarily have a merger origin. 65\% of HST-SPOGs have a higher shape asymmetry than their comparison galaxies, likely caused by tidal features induced during a merger. However, 30\% of HST-SPOGs with high asymmetry do not have a high shape asymmetry, and instead show internal disturbances, likely caused by internal disturbances or dusty structures. Internal and dust morphology could be affected by a range of instabilities: outflows, bars, turbulence, mergers, or a combination thereof. 



It is possible that most or all of the galaxies in our sample experienced a merger, which triggered the instabilities we see now. The similarity in asymmetry timescales between our sample and merger simulations \citep[e.g.,][]{Pawlik2018, Snyder2015a} suggests a possible merger origin. However, obvious merger signatures (such as tidal tails) may have already faded, so confirming a previous merger event is difficult. To fully understand the structure and the origin of the internal disturbances observed in our sample, we plan to follow up this study with a larger \textit{HST} survey of the parent SPOG sample, spanning a range of masses, redshifts, post-starburst ages, and environments. In addition, deeper \textit{HST} imaging or the next-generation deep imaging from the Vera Rubin Observatory will allow detecting the faintest tidal features in post-starburst galaxies, and therefore constraining their merger origin. The new high-resolution simulations of post-starbursts with a full time evolution information will also help to determine if they their starbursts are triggered by a merger  \citep[e.g.,][]{Zheng2020, Lotz2020}. 

\acknowledgements

We thank the anonymous referee for the thoughtful and constructive feedback they provided on this manuscript, helping to strengthen the discussion of our results. ES thanks Mark Lacy, Roan Haggar, and Nathan Miles for very useful scientific suggestions and feedback. ES, KA, and YL have been partially funded by Space Telescope Science Institute Director's Discretionary Research Fund grants D0101.90241 and D0101.90262, and \textit{HST} grants GO-14715.021, GO-14649.015, and \textit{Chandra} grant GO7-18096A. AMM acknowledges support by the National Science Foundation under Grant No. 2009416. UL acknowledge support by the research project  AYA2017-84897-P from the Spanish Ministerio de Economía y Competitividad, from the European Regional Development Funds (FEDER) and the Junta de Andalucía (Spain) grant FQM108. 

This research made use of the following Python astronomy libraries: Astropy, a community-developed core Python package for Astronomy \citep{astropy,astropy2}; \textsc{statmorph}, open-source package for morphological measurements \citep{statmorph}; Photutils,  an Astropy package for detection and photometry of astronomical sources \citep{photutils}; Astro-SCRAPPY, an Astropy package for detection of cosmic rays \citep{astroscrappy}; ccdproc, an Astropy package for
image reduction \citep{ccdproc}. This work also used \textsc{GALFIT}, a software for galaxy profile fitting \citep{galfit,galfit2}, and TinyTim, a software to model the PSF of \textit{HST} images \citep{Krist2011}.

This research made of data products from the Slon Digital Sky Survey and the NASA/ESA \textit{Hubble Space Telescope}. \textit{HST} observations were obtained from the Hubble Legacy Archive, which is a collaboration between the Space Telescope Science Institute (STScI/NASA), the Space Telescope European Coordinating Facility (ST-ECF/ESA) and the Canadian Astronomy Data Centre (CADC/NRC/CSA). Funding for the Sloan Digital Sky Survey IV has been provided by the Alfred P. Sloan Foundation, the U.S. Department of Energy Office of Science, and the Participating Institutions. 

SDSS-IV acknowledges support and resources from the Center for High Performance Computing  at the University of Utah. The SDSS website is www.sdss.org.SDSS-IV is managed by the Astrophysical Research Consortium for the Participating Institutions of the SDSS Collaboration including the Brazilian Participation Group, the Carnegie Institution for Science, Carnegie Mellon University, Center for Astrophysics | Harvard \& Smithsonian, the Chilean Participation Group, the French Participation Group, Instituto de Astrof\'isica de Canarias, The Johns Hopkins University, Kavli Institute for the Physics and Mathematics of the Universe (IPMU) / University of Tokyo, the Korean Participation Group, Lawrence Berkeley National Laboratory, Leibniz Institut f\"ur Astrophysik Potsdam (AIP),  Max-Planck-Institut f\"ur Astronomie (MPIA Heidelberg), Max-Planck-Institut f\"ur Astrophysik (MPA Garching), Max-Planck-Institut f\"ur Extraterrestrische Physik (MPE), National Astronomical Observatories of China, New Mexico State University, New York University, University of Notre Dame, Observat\'ario Nacional / MCTI, The Ohio State University, Pennsylvania State University, Shanghai Astronomical Observatory, United Kingdom Participation Group, Universidad Nacional Aut\'onoma de M\'exico, University of Arizona, University of Colorado Boulder, University of Oxford, University of Portsmouth, University of Utah, University of Virginia, University of Washington, University of Wisconsin, Vanderbilt University, and Yale University.

\software{Astropy \citep{astropy,astropy2}, \textsc{statmorph} \citep{statmorph}, \textsc{GALFIT} \citep{galfit,galfit2}, TinyTim \citep{Krist2011}, photutils \citep{photutils}, Astro-SCRAPPY \citep{astroscrappy}, ccdproc \citep{ccdproc}, pyQt-Fit\footnote{\href{https://pyqt-fit.readthedocs.io}{PyQt-Fit webpage - pyqt-fit.readthedocs.io}}, Pandas \citep{pandas}, SciPy \citep{scipy}, Matplotlib \citep{matplotlib}, NumPy \citep{numpy}.}


\bibliographystyle{mnras_custom}
\raggedright

\vspace{-4mm}
\setlength{\bibsep}{0mm}
\bibliography{references}

\appendix 


\section{Morphological parameters of each galaxy in the sample}\label{app:morphology}

Tab. \ref{tab:all_morphology} shows a sample of a few morphological measurements outlined in Sec. \ref{sec:morphology} for a few galaxies. The entire table showing all of the measurements for each galaxy and each imaging filter is available in the machine-readable format in the online journal.

\section{Statistics and distributions of all morphological parameters}\label{app:more_stats}

Tab. \ref{tab:medians_all} shows the medians of all morphological parameters for the HST-SPOGs, star-forming and quiescent samples. Same as in Tab. \ref{tab:medians_filters} and \ref{tab:medians}, the uncertainty on the median is calculated by bootstrapping each sample 1,000 times, and finding the 16\ts{th}/84\ts{th} quantiles of the median distribution. The same general trends can be seen as described in Sec. \ref{sec:filt_morph} and Sec. \ref{sec:populations}. HST-SPOGs have intermediate bulge strengths between star-forming and quiescent galaxies in all 5 structural parameters. They also have a higher bulge strength and more compact morphology in redder imaging than bluer imaging. The only exception is $B/T$ -- $B/T$ measured with F438W and F814W imaging are consistent with each other. HST-SPOGs are significantly more disturbed than comparison galaxies in \textit{HST} imaging, and the disturbance increases in bluer imaging, likely an effect of dust substructure as discussed in Sec. \ref{sec:a_violins}.

\begin{figure*}
    \centering
    \includegraphics[width=\linewidth]{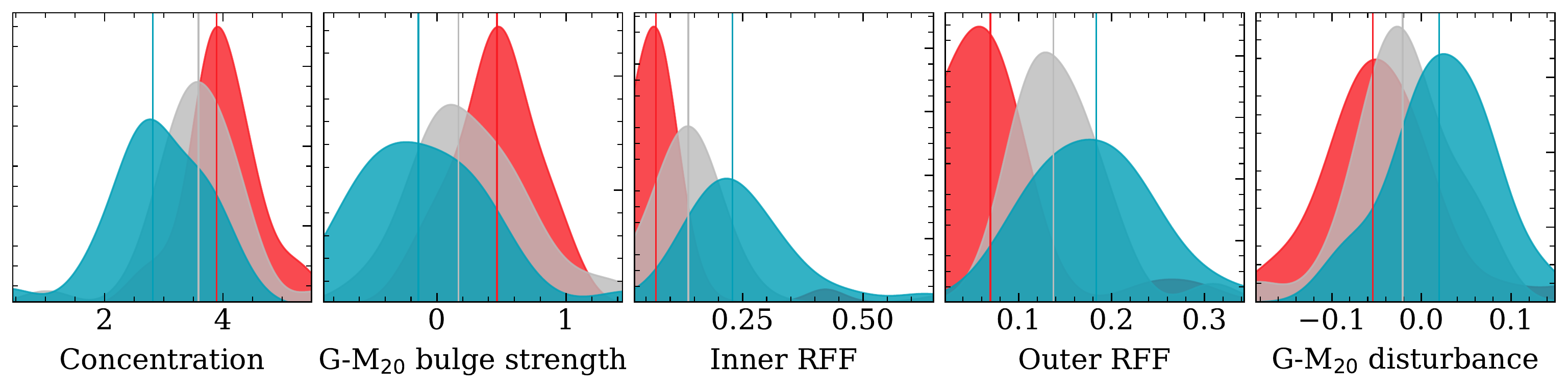}
    \caption{Distributions of morphological parameters from Sec. \ref{sec:morphology} not given in Fig. \ref{fig:filter_dist} for the HST-SPOG sample obtained with \textit{HST} F160W (red), F814W (grey) and F438W (blue) imaging. Left to right, the parameters are: concentration, G-M$_{20}$ bulge strength, inner RFF, outer RFF, and G-M$_{20}$ disturbance. HST-SPOGs have stronger bulges in redder imaging, and are more disturbed in bluer imaging.}
    \label{fig:filt_morphs_all}
\end{figure*}

\begin{figure*}
    \centering
    \includegraphics[width=\linewidth]{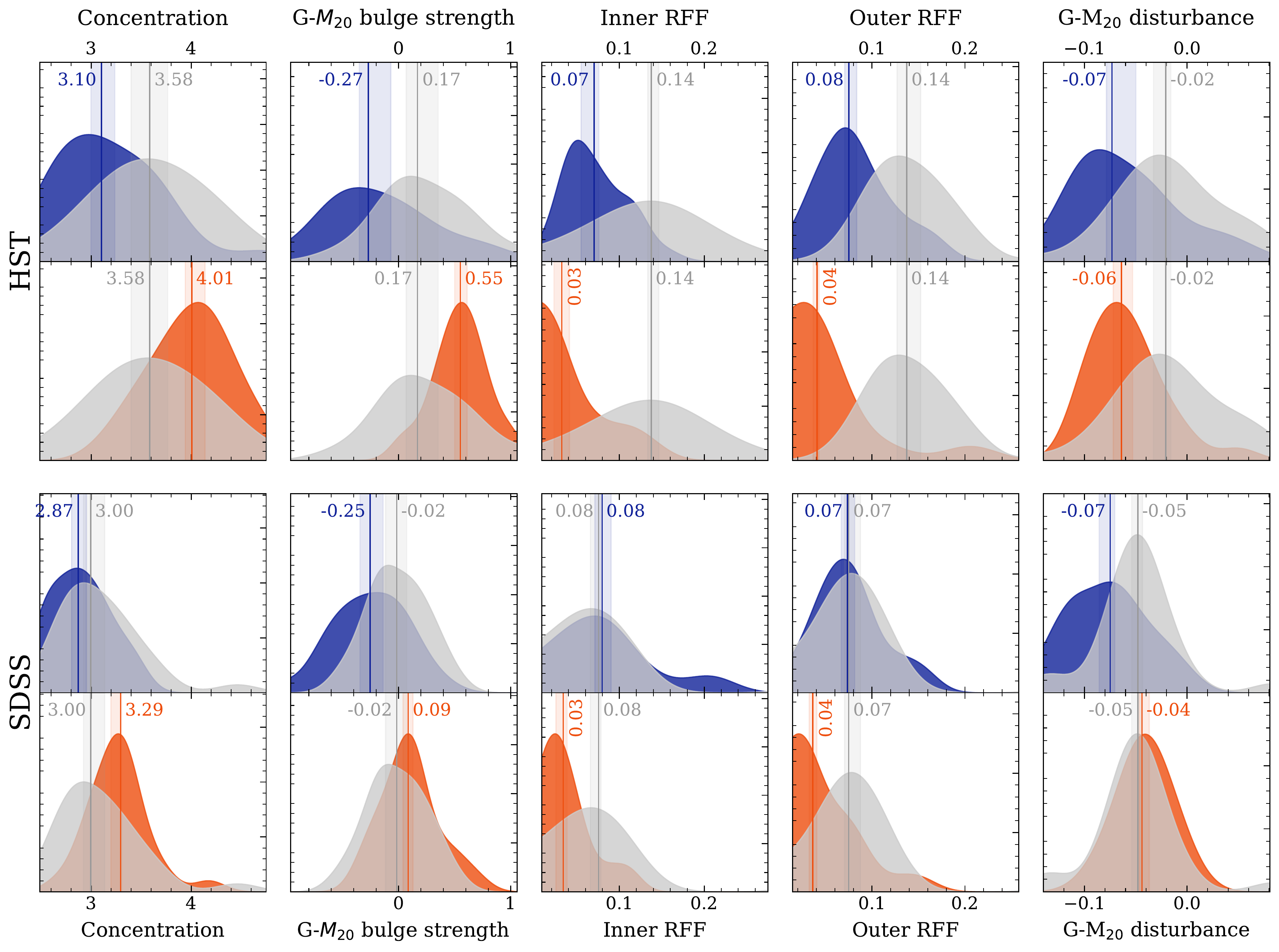}
    \caption{Distributions of the same morphological parameters as in Fig. \ref{fig:filt_morphs_all} for HST-SPOGs (grey),  star-forming galaxies (blue) and quiescent galaxies (orange). Parameters were computed using \textit{HST} F814W imaging (top) and SDSS \textit{i}-band imaging (bottom). HST-SPOGs have intermediate bulge strengths between star-forming and quiescent galaxies, and are significantly more disturbed when observed with \textit{HST}.}
    \label{fig:morphs_all}
\end{figure*}

Fig. \ref{fig:filt_morphs_all} and \ref{fig:morphs_all} show the distributions of morphological parameters from Sec. \ref{sec:morphology} that were not displayed in Fig. \ref{fig:filter_dist} and \ref{fig:morph_distributions}, respectively. Fig. \ref{fig:filt_morphs_all} shows a stronger dependence of morphological parameters on wavelength than Fig. \ref{fig:filter_dist}, indicating that concentration and G-M$_{20}$ metrics are more wavelength-dependent than Sérsic index, $B/T$ and asymmetry. The general trends are the same: HST-SPOGs are more disturbed in bluer bands and more disturbed than comparison galaxies using all metrics. They have intermediate bulge strengths between star-forming and quiescent galaxies, and redder central bulges.

\begin{deluxetable*}{chhhhhhh|ccchh|ccchhhc|cchcch}
\label{tab:all_morphology}
\tablecaption{Morphological measurements for HST-SPOGs, star-forming and quiescent galaxies calculated in this work (sample).}
\tablehead{
\multicolumn{8}{c}{Galaxy properties} & \multicolumn{5}{c}{Image properties} & \multicolumn{7}{c}{Structural parameters} & \multicolumn{6}{c}{Disturbance parameters} \\
IAU Name & Sample & R.A. & Dec. & log M$_\star/$M$_\odot$ & Redshift & Comparison QG & Comparison SFG & Instrument & Filter & Depth & PSF FWHM & Pixel scale & R$_p$ & Sérsic n & B/T & G & M$_{20}$ & G-M$_{20}$ bulge strength & $C$ & $A$ & $A_S$ & G-M$_{20}$ Dist. & RFF$_{in}$ & RFF & RFF$_{out}$}
\startdata
J000318+004844 & HST-SPOG & 0.826 & 0.812 & 10.820 & 0.139 & J095853+022603 & J162904+470853 & HST & F160W & 23.401 & 0.153 & 0.128 & 4.509 & 2.837 & 0.799 & 0.591 & -1.946 & 0.314 & 3.886 & 0.196 & 0.472 & -0.013 & 0.066 & 0.050 & 0.043 \\
J000318+004844 & HST-SPOG & 0.826 & 0.812 & 10.820 & 0.139 & J095853+022603 & J162904+470853 & HST & F438W & 23.439 & 0.070 & 0.040 & 4.957 & 1.000 & 0.154 & 0.555 & -1.198 & -0.381 & 2.759 & 0.010 & 0.424 & 0.056 & 0.241 & 0.201 & 0.169 \\
J000318+004844 & HST-SPOG & 0.826 & 0.812 & 10.820 & 0.139 & J095853+022603 & J162904+470853 & HST & F814W & 22.252 & 0.074 & 0.040 & 5.624 & 1.646 & 0.468 & 0.549 & -1.549 & -0.170 & 3.161 & 0.010 & 0.528 & 0.001 & 0.172 & 0.163 & 0.154 \\
J000318+004844 & HST-SPOG & 0.826 & 0.812 & 10.820 & 0.139 & J095853+022603 & J162904+470853 & SDSS & i & 23.713 & 0.906 & 0.396 & 6.752 & 0.954 & 0.000 & 0.494 & -1.596 & -0.411 & 2.780 & 0.011 & 0.359 & -0.060 & 0.075 & 0.089 & 0.112 \\
J001145-005431 & HST-SPOG & 2.938 & -0.909 & 10.222 & 0.048 & J160210+155639 & J123704+260525 & HST & F160W & 23.397 & 0.153 & 0.128 & 2.719 & 2.044 & 0.424 & 0.595 & -2.278 & 0.561 & 3.784 & 0.023 & 0.142 & -0.055 & 0.071 & 0.034 & 0.021 \\
J001145-005431 & HST-SPOG & 2.938 & -0.909 & 10.222 & 0.048 & J160210+155639 & J123704+260525 & HST & F438W & 23.557 & 0.070 & 0.040 & 3.067 & 0.944 & 1.000 & 0.542 & -1.461 & -0.266 & 2.816 & 0.084 & 0.311 & 0.006 & 0.197 & 0.157 & 0.125 \\
J001145-005431 & HST-SPOG & 2.938 & -0.909 & 10.222 & 0.048 & J160210+155639 & J123704+260525 & HST & F814W & 22.405 & 0.074 & 0.040 & 2.883 & 1.595 & 0.321 & 0.584 & -1.973 & 0.299 & 3.347 & 0.054 & 0.319 & -0.023 & 0.119 & 0.109 & 0.100 \\
J001145-005431 & HST-SPOG & 2.938 & -0.909 & 10.222 & 0.048 & J160210+155639 & J123704+260525 & SDSS & i & 23.630 & 0.906 & 0.396 & 3.045 & 1.173 & 0.121 & 0.554 & -1.903 & 0.098 & 3.124 & -0.067 & 0.154 & -0.043 & 0.068 & 0.065 & 0.058 \\
\enddata
\tablecomments{A subset of morphological measurements obtained with \textit{HST} F160W, F814W, F438W and SDSS \textit{i}-band imaging of two sample HST-SPOGs. Full dataset for all galaxies used in this work and all parameters is given as a machine-readable table in the online journal. (1): IAU name. (3) and (4): instrument and filter used for the measurement. (5): 1$\sigma$ sky background flux of the image in AB mag/arcsec\ts{2}. Structural parameters: (6) Petrosian radius, (7) Sérsic index, (8) bulge-to-total light ratio. Disturbance parameters: (9) asymmetry, (10) shape asymmetry, and (12) residual flux fraction. Additional data for all galaxies and all morphological parameters is available in the machine-readable table in the online journal.}
\end{deluxetable*}

\renewcommand{\arraystretch}{1.2}
\begin{deluxetable*}{r|D|D|DDD|DDD}
\label{tab:medians_all}
\tablecaption{Medians of all morphological measurements for HST-SPOGs and comparison star-forming and quiescent galaxies, calculated with \textit{HST} F160W, F438W, F814W imaging (where available) and SDSS \textit{i}-band imaging.}
\tablehead{
\multicolumn1r{} & \multicolumn2c{\textit{HST} F160W} & \multicolumn2c{\textit{HST} F438W} & \multicolumn6c{\textit{HST} F814W} & \multicolumn6c{SDSS i-band} \\
\multicolumn1r{Parameter} & \multicolumn2c{HST-SPOGs} & \multicolumn2c{HST-SPOGs} & \multicolumn2c{HST-SPOGs} & \multicolumn2c{SFGs} & \multicolumn2c{QGs} & \multicolumn2c{HST-SPOGs} & \multicolumn2c{SFGs} & \multicolumn2c{QGs}
}
\decimals
\startdata
\multicolumn{17}{l}{Structural parameters} \\
\hline
Sérsic $n$ & 1.98$^{+0.35}_{-0.08}$ & 1.18$^{+0.41}_{-0.18}$ & 1.70$^{+0.19}_{-0.10}$ & 1.22$^{+0.23}_{-0.12}$ & 2.64$^{+0.34}_{-0.24}$ & 1.43$^{+0.05}_{-0.06}$ & 0.87$^{+0.18}_{-0.08}$ & 2.04$^{+0.03}_{-0.11}$ \\
$B/T$ & 0.64$^{+0.07}_{-0.05}$ & 0.57$^{+0.09}_{-0.22}$ & 0.47$^{+0.05}_{-0.10}$ & 0.16$^{+0.01}_{-0.01}$ & 0.71$^{+0.06}_{-0.05}$ & 0.17$^{+0.03}_{-0.03}$ & 0.03$^{+0.04}_{-0.01}$ & 0.38$^{+0.02}_{-0.03}$ \\
Concentration & 3.89$^{+0.18}_{-0.09}$ & 2.82$^{+0.21}_{-0.08}$ & 3.66$^{+0.13}_{-0.21}$ & 3.10$^{+0.13}_{-0.11}$ & 4.01$^{+0.13}_{-0.07}$ & 3.00$^{+0.13}_{-0.07}$ & 2.87$^{+0.10}_{-0.07}$ & 3.29$^{+0.01}_{-0.10}$ \\
G-M$_{20}$ bulge strength & 0.47$^{+0.07}_{-0.07}$ & -0.14$^{+0.10}_{-0.12}$ & 0.27$^{+0.12}_{-0.20}$ & -0.27$^{+0.19}_{-0.08}$ & 0.55$^{+0.06}_{-0.07}$ & -0.02$^{+0.08}_{-0.11}$ & -0.25$^{+0.12}_{-0.09}$ & 0.09$^{+0.04}_{-0.05}$ \\
R$_{0.5}$ [kpc] & 2.92$^{+0.15}_{-0.48}$ & 3.06$^{+0.46}_{-0.30}$ & 3.21$^{+0.39}_{-0.33}$ & 4.39$^{+0.55}_{-0.35}$ & 2.39$^{+0.22}_{-0.20}$ & 3.92$^{+0.63}_{-0.23}$ & 5.47$^{+0.36}_{-0.39}$ & 3.43$^{+0.26}_{-0.32}$ \\
\hline
\multicolumn{17}{l}{Disturbance parameters} \\
\hline
Asymmetry & 0.09$^{+0.04}_{-0.01}$ & 0.12$^{+0.07}_{-0.02}$ & 0.13$^{+0.01}_{-0.03}$ & -0.03$^{+0.01}_{-0.01}$ & -0.01$^{+0.01}_{-0.01}$ & 0.03$^{+0.01}_{-0.01}$ & -0.00$^{+0.02}_{-0.02}$ & -0.00$^{+0.01}_{-0.00}$ \\
Shape asymmetry & 0.29$^{+0.07}_{-0.04}$ & 0.39$^{+0.04}_{-0.04}$ & 0.30$^{+0.03}_{-0.03}$ & 0.27$^{+0.01}_{-0.02}$ & 0.20$^{+0.01}_{-0.01}$ & 0.27$^{+0.02}_{-0.04}$ & 0.30$^{+0.03}_{-0.02}$ & 0.22$^{+0.02}_{-0.01}$ \\
RFF & 0.07$^{+0.00}_{-0.01}$ & 0.22$^{+0.00}_{-0.01}$ & 0.14$^{+0.01}_{-0.01}$ & 0.07$^{+0.01}_{-0.01}$ & 0.04$^{+0.01}_{-0.01}$ & 0.07$^{+0.01}_{-0.01}$ & 0.07$^{+0.01}_{-0.01}$ & 0.03$^{+0.01}_{-0.00}$ \\
RFF\sub{in} & 0.07$^{+0.00}_{-0.00}$ & 0.23$^{+0.02}_{-0.02}$ & 0.14$^{+0.01}_{-0.00}$ & 0.07$^{+0.01}_{-0.02}$ & 0.03$^{+0.01}_{-0.01}$ & 0.08$^{+0.00}_{-0.01}$ & 0.08$^{+0.01}_{-0.01}$ & 0.03$^{+0.00}_{-0.01}$ \\
RFF\sub{out} & 0.07$^{+0.00}_{-0.02}$ & 0.19$^{+0.02}_{-0.02}$ & 0.14$^{+0.01}_{-0.02}$ & 0.08$^{+0.01}_{-0.00}$ & 0.04$^{+0.00}_{-0.00}$ & 0.07$^{+0.01}_{-0.00}$ & 0.07$^{+0.01}_{-0.01}$ & 0.04$^{+0.00}_{-0.00}$ \\
G-M$_{20}$ disturbance & -0.05$^{+0.02}_{-0.01}$ & 0.02$^{+0.03}_{-0.01}$ & -0.02$^{+0.01}_{-0.01}$ & -0.07$^{+0.02}_{-0.01}$ & -0.06$^{+0.01}_{-0.01}$ & -0.05$^{+0.01}_{-0.01}$ & -0.07$^{+0.00}_{-0.01}$ & -0.04$^{+0.01}_{-0.00}$ \\
\enddata
\end{deluxetable*}

\end{document}